\documentclass[10pt,journal]{IEEEtran}
\usepackage{amsmath,amsfonts}
\usepackage{array}
\usepackage[caption=false,font=footnotesize,labelfont=rm,textfont=rm]{subfig}
\usepackage{textcomp}
\usepackage{stfloats}
\usepackage{url}
\usepackage{bm}
\usepackage{verbatim}
\usepackage{graphicx}
\usepackage{multirow}
\usepackage{booktabs}
\usepackage{color}
\usepackage{soul}
\usepackage{hyperref}
\usepackage[ruled,linesnumbered]{algorithm2e}
\def\BibTeX{{\rm B\kern-.05em{\sc i\kern-.025em b}\kern-.08em
    T\kern-.1667em\lower.7ex\hbox{E}\kern-.125emX}}
\usepackage{balance}

\hypersetup{
    colorlinks=true,  
    linkcolor=black,  
    citecolor=green,  
    urlcolor=black     
}

\begin{document}
\title{UAV Cognitive Semantic Communications \\ Enabled by Knowledge Graph for Robust Object Detection}
\author{
  Xi Song, Fuhui Zhou, \textit{Senior Member}, \textit{IEEE}, Rui Ding, Zhibo Qu, Yihao Li, Qihui Wu, \textit{Fellow}, \textit{IEEE}, \\
  and Naofal Al-Dhahir, \textit{Fellow}, \textit{IEEE}
\thanks{

  This work was presented in part at the IEEE International Conference on Communications (ICC) \cite{song2024knowledge}, Denver, CO, USA, Jun. 2024.

  This work was supported in part by the National Key R\&D Program of China under Grant 2023YFB2904500; in part by the National Natural Science Foundation of China under Grant 62222107; in part by the Basic Reserch Projects of Stabilizing Support for Specialty Disciplines under Grant No.ILF240041A24; in part by the Yangtze River Delta Science and Technology Innovation Community Joint Research (Basic Research) Project under Grant 1030-PQB24004. The work of N. Al-Dhahir was supported by Erik Jonsson Distinguished Professorship.

  X. Song, R. Ding, Z. Qu, Y. Li and Q. Wu are with the College of Electronic and Information Engineering, Nanjing University of Aeronautics and Astronautics, P. R. China, 330031 (e-mail: sx32510@nuaa.edu.cn; rui\_ding@nuaa.edu.cn; zhiboqu@nuaa.edu.cn; liyihao1999@nuaa.edu.cn; wuqihui2014@sina.com).

  F. Zhou is with the College of Artificial Intelligence, Nanjing University of Aeronautics and Astronautics, P. R. China, 330031 (e-mail: zhoufuhui@ieee.org).

  N. Al-Dhahir is with the Department of Electrical and Computer Engineering, The University of Texas at Dallas, Richardson, TX 75080 USA (e-mail: aldhahir@utdallas.edu).

  The corresponding author is Fuhui Zhou.
  
  }
}
\maketitle

\begin{abstract}
  Unmanned aerial vehicles (UAVs) are widely used for object detection. However, the existing UAV-based object detection systems are subject to severe challenges, namely, their limited computation, energy and communication resources, which limits the achievable detection performance. To overcome these challenges, a UAV cognitive semantic communication system is proposed by exploiting a knowledge graph. Moreover, we design a multi-scale codec for semantic compression to reduce data transmission volume while guaranteeing detection performance. Considering the complexity and dynamicity of UAV communication scenarios, a signal-to-noise ratio (SNR) adaptive module with robust channel adaptation capability is introduced. Furthermore, an object detection scheme is proposed by exploiting the knowledge graph to overcome channel noise interference and compression distortion. Simulation results conducted on the practical aerial image dataset demonstrate that our proposed semantic communication system outperforms benchmark systems in terms of detection accuracy, communication robustness, and computation efficiency, especially in dealing with low bandwidth compression ratios and low SNR regimes.
\end{abstract}


\section{Introduction}

  \subsection{Background and Motivations}
  \IEEEPARstart{D}{ue} to the advantages of flexible deployment, low cost and versatility, unmanned aerial vehicles (UAVs) are used for object detection in various civil and military applications, such as military reconnaissance, environmental monitoring, and urban planning \cite{Bouguettaya2022vehicle}. Object detection plays an increasingly important role in the missions carried out by UAVs \cite{li2023cross}. However, existing UAV object detection systems are subject to severe challenges such as limited endurance time and high computation latency because of the finite battery capacity, computation and communication resources \cite{ding2024external}. Therefore, it is imperative to develop innovative UAV-based object detection systems to overcome these challenges.

  The existing UAV-based object detection frameworks can be primarily classified into two categories. The first one is to deploy the object detection algorithms on the UAV, which utilizes the computation resources, including processing units, memory, and sensors, to perform real-time detection tasks autonomously \cite{lai2023realtime}. A key advantage of this framework is its independence from external infrastructure, making it particularly well-suited for scenarios that demand rapid response times. Moreover, the first framework does not require data transmission and the consumption of spectrum bandwidth. The second framework is a UAV-server collaboration framework. In this framework, UAVs collect raw sensor data and transmit them to the remote computing servers for processing and analysis \cite{zhao2022multi}. This framework leverages the computation resources of the servers to execute computationally intensive detection algorithms to overcome the limitations of onboard processing capabilities. Moreover, lightweight UAVs with reduced onboard computation payloads can be achieved by offloading processing tasks to the computing servers. Consequently, the endurance and operational range can be extended in the second framework \cite{ji2021joint}.

  However, both frameworks encounter challenges that require to be addressed urgently. The first framework cannot be widely used due to the UAV's finite computation capabilities and power constraints, potentially compromising the detection accuracy and scalability for complex tasks or large-scale deployments. The reliance on robust connections between UAVs and servers makes the second framework susceptible to latency, bandwidth, and reliability issues \cite{haber2021uav}.

  Fortunately, semantic communication is promising in addressing the above-mentioned challenges, which can realize an excellent trade-off between detection accuracy and required energy, computation and communication resources. It was defined by Weaver and Shannon \cite{xu2012opportunistic}. In contrast to conventional wireless communication systems prioritizing the successful transmission of symbols between the transmitter and the receiver, semantic communication transmits the semantic information at the transmitter and interprets the semantic meaning at the receiver. The data transmission volume can be reduced significantly, and the system robustness at low signal-to-noise ratios (SNRs) can be enhanced remarkably \cite{qin2021semantic}.

  Semantic communication can be broadly categorized into three paradigms, namely, compression-recovery semantic communication \cite{kurka2021bandwidth}, task-oriented semantic communication \cite{kang2022task}, and knowledge-based semantic communication \cite{yi2023deep}. Compression-recovery semantic communication aims to recover the raw data at the receiver through semantic features compressed at the transmitter \cite{erdemir2023generative}. For the task-oriented semantic communication, task-related semantic features take precedence. Therefore, this paradigm is driven by specific tasks and metrics with semantic extraction focused on capturing semantic features to enhance the task performance \cite{zeng2024task}. The knowledge-based semantic communication achieves semantic compression and communication with the assistance of external knowledge bases such as knowledge graph (KG) \cite{zhou2023cognitive}.

  Current research mainly focuses on the first two paradigms \cite{wang2023knowledge}. Although the first two paradigms can alleviate the bandwidth limitation and the noise interference problems in object detection within UAV communication scenarios, they struggle to further address issues such as high resource consumption, poor real-time performance, and low accuracy. The main characteristics of knowledge-based semantic communication are that it achieves learning, understanding, reasoning, and adaptive error correction with the aid of external knowledge. Hence, introducing external knowledge can achieve higher accuracy from limited information when the UAV communication environment is poor. Additionally, by supplementing semantic information with external knowledge, a more lightweight network can be used during semantic extraction, thus improving real-time performance and reducing resource consumption.

  Motivated by the above-mentioned facts, a UAV cognitive semantic communication system for object detection is developed by exploiting a knowledge graph in this paper. The system utilizes a UAV-server collaboration framework, which dramatically reduces the resource usage of the UAV. In contrast to the traditional UAV-server collaboration framework, cognitive semantic communication with a knowledge graph is introduced to counteract the inevitable compression distortion and noise interference. Specifically, the correlation and similarity between entities of the knowledge graph are exploited to enhance the global semantic understanding of the proposed system. Errors in received data typically arise from both the compression and transmission phases in the communication process. During the compression phase, a multi-scale codec is developed to compress the semantic features in multiple granularities to decrease the bandwidth compression ratio while ensuring fidelity. Noise interference in UAV communication scenarios possesses complexity and dynamicity, which is undoubtedly a significant challenge. An SNR adaptive (SA) module is incorporated to enable the proposed system to be adaptive to interference across a wide range of SNRs.

  Extensive simulations are performed on a practical aerial image dataset to demonstrate the effectiveness of our system. The conventional method and an attention deep learning-based joint source and channel method (ADJSCC) are used as benchmarks for comparison \cite{xu2021wireless}. ADJSCC is an image semantic communication method that utilizes convolutional neural networks and channel attention mechanisms for joint source-channel encoding and decoding of images. Unlike our system, both benchmarks reconstruct lossy images at the receiver. We assume that benchmarks operate within a UAV-server collaborative framework. The transmission modules are deployed on the UAV, while the reception and detection modules are deployed on the server. To ensure a fair comparison, the detection task is performed using Faster R-CNN on the benchmarks at the receiver, while our system carries out the detection task with an identically configured detector.

  Object detection is adopted as the task scenario to demonstrate the generality of our framework. On the one hand, object detection involves both classification and regression. On the other hand, object detection serves as an upstream support for many tasks in computer vision. Therefore, our framework can be generalized to various tasks, such as image classification, object tracking, visual question answering, anomaly detection, etc., by simply constructing the corresponding knowledge graphs for each domain.

  \subsection{Related Work}

  \textbf{UAV-based Object Detection.} UAV-based object detection has been a popular research area in recent years. Within the UAV-server collaboration framework, the majority of works were dedicated to offloading the object detection task to computing servers. The authors in \cite{lee2017real} proposed moving the computation to an off-board computing server while keeping low-level object detection and short-term navigation onboard. Moreover, faster regions with convolutional neural networks are applied to detect hundreds of object types in near real-time. In \cite{ozer2023offloading}, the authors designed an object detection system to transmit deep learning-powered vision tasks from the UAV to a computing server. In the UAV-server collaboration framework, the hardware devices change while the scenario varies. In \cite{wang2022online}, the UAVs in the network are classified into a mission UAV for collecting images and a computing UAV equipped with an edge server for detecting objects according to their computing capabilities. The task allocation algorithm is at the centre of the UAV-server collaboration framework. To facilitate adaptive task allocation between the edge-side and cloud-side, the authors in \cite{yuan2024edgecloud} presented a distributed edge-cloud collaborative framework for UAV object detection through a decision-making mechanism based on a fuzzy neural network.

  \textbf{Semantic Communication.} Compression-recovery semantic communication is the most extensively researched among the three paradigms \cite{luo2022semantic}. Inspired by deep learning-based image compression methods, the authors in \cite{bourtsoulatze2019deepjscc} presented a deep learning-based joint source and channel coding technique for image transmission. The encoder and decoder functions were parameterized by two convolutional neural networks and trained jointly. The overall network was considered an autoencoder with a non-trainable layer in the middle representing the noisy communication channel. However, the network was trained to operate under a specific SNR regime without taking into account the variability of communication conditions. To tackle this problem, the authors in \cite{xu2021wireless} proposed a method called attention deep learning-based joint source and channel coding. This method successfully operated with different SNR regimes during transmission. Besides image wireless transmission, semantic communication systems for other modalities have been sequentially investigated \cite{peng2024arobust}, \cite{han2023semantic}. A robust text semantic communication system was proposed by introducing a semantic corrector for robust semantic encoding \cite{peng2024arobust}. For the transmission of speech, attention-based semantic communication was developed to recover speech signals at the receiver \cite{han2023semantic}.

  Task-oriented semantic communication addresses the limitations of the first paradigm in specific task scenarios. In these scenarios, the semantic features recovered by the receiver are unable to align with those required for the tasks \cite{ma2023task}. Task-oriented semantic communication extracts the most relevant semantic information within the source data for task execution at the receiver. Filtering out task-irrelevant semantic features reduces both bandwidth consumption and transmission latency \cite{gunduz2022beyond}. There have also been many research efforts on task-oriented semantic communication for vision tasks. In \cite{xie2022task}, the authors proposed three Transformer-based models for image retrieval, machine translation, and visual question answering tasks. In \cite{zhang2022multi}, a multi-user semantic communication system was studied to execute object detection tasks, where correlated source data among different users is transmitted via a shared channel. The authors in \cite{wu2022semantic} introduced transfer learning to improve the accuracy of the semantic communication system for few-shot object detection. In \cite{fu2023content}, the authors developed a content-aware semantic communication system for both the image reconstruction task and the object detection task. 

  Knowledge-based semantic communication is still in the early research stage and is rarely studied since it requires interdisciplinary knowledge \cite{yang2023semantic}. The authors in \cite{emilio2021semantic} presented a survey and suggested that the knowledge graph can serve as a sharing knowledge system to realize semantic communication. However, the specific system and implementation were not presented. The concept of ``cognitive semantic communication'' was first presented in \cite{zhou2023cognitive}, which exploited a knowledge graph to achieve semantic compression and semantic inference to improve transmission efficiency and accuracy. The authors demonstrated that the exploitation of the knowledge graph significantly improved communication efficiency and reliability and proposed two cognitive semantic communication systems for single-user and multi-user semantic communication scenarios.
  
  \subsection{Contributions and Organization}

  Our work in this paper is the extension of our previous work in \cite{song2024knowledge}. The dynamic and time-varying characteristics of noise interference in UAV communication scenarios are further considered. An SA module with robust channel adaptation capability is designed for our proposed system. Moreover, extensive simulation is conducted to validate the robustness of the system. The main contributions of this paper are summarized as follows.

  \begin{itemize}
    \item{To address the challenge of constrained resources in UAV-based object detection, a novel UAV object detection system is designed driven by cognitive semantic communication. The knowledge graph is exploited to develop the detection scheme. The auxiliary knowledge from the knowledge graph establishes associations with semantic features and synergizes the detection to enhance accuracy.}
    \item{Two methods are proposed to enhance the stability and robustness of the semantic communication system, especially in the context of UAV applications. To mitigate compression distortion, a multi-scale codec is proposed by leveraging a multi-scale approach to encode and decode semantic features. To counteract the effects of dynamic noise interference, we present a strategically designed SA module to ensure reliable communication in challenging UAV scenarios.}
    \item{Extensive simulation is performed on a practical aerial image dataset. It is demonstrated that our proposed system outperforms benchmark systems in terms of detection accuracy, communication robustness, and computation efficiency, especially in dealing with low bandwidth compression ratios and low SNR regimes. The effectiveness of the designed multi-scale codec and SA module is further demonstrated through the ablation study.}
  \end{itemize}

  The remainder of this paper is organized as follows. Section II presents the UAV cognitive semantic communication system. Section III elaborates on the details of our proposed system for the implementation. Simulation results are provided in Section IV. Finally, the paper concludes with Section V.

  \begin{figure*}[!t]
    \centering
    \includegraphics[width=6.8 in]{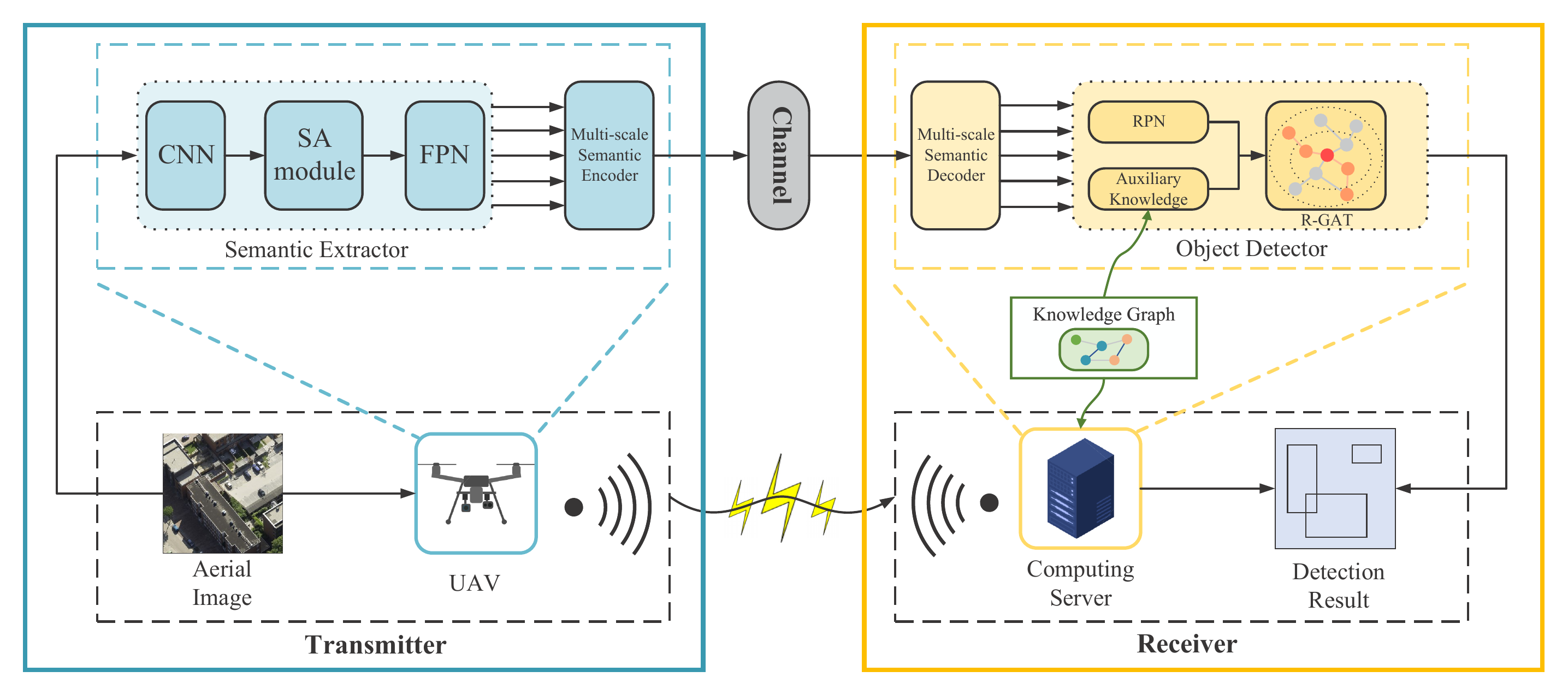}
    \caption{The proposed UAV cognitive semantic communication system.}
    \label{framework}
  \end{figure*}

\section{UAV Cognitive Semantic Communication \\ For Object Detection}
  In this section, a UAV cognitive semantic communication system is proposed, which consists of three modules, namely, a semantic extractor, a multi-scale semantic codec, and an object detector are included. The semantic extractor and the multi-scale semantic encoder are deployed on a computation-resource-constrained UAV, whereas the multi-scale semantic decoder and the object detector are implemented on a computing server with sufficient computation resources. Moreover, a knowledge graph is exploited in our proposed UAV cognitive semantic communication system to enhance detection performance and communication robustness. The specific architecture of our proposed system is shown in Fig. \ref{framework}.

  Inspired by recent computer vision development, we exploit an innovative semantic extractor for the UAV semantic communication system to extract the semantic features of aerial images. To alleviate the computation resource consumption, a lightweight convolutional neural network is applied as the backbone of the semantic extractor. To tackle the challenge of the multi-scale variation prevalent in aerial images, the feature pyramid network (FPN) is incorporated into the semantic extractor \cite{liao2023realtime}. The FPN is an architecture used in convolutional neural networks to enhance object detection by efficiently merging information across different scales. It generates a multi-scale feature pyramid by combining high-resolution, low-level features with low-resolution, high-level features through top-down pathways and lateral connections. This allows the model to detect objects of varying sizes more effectively.

  The aim of semantic encoding and decoding is to reduce semantic redundancy while compressing semantic features. Compared to other semantic communication systems, our semantic encoder and decoder adopt a novel parallel architecture. Specifically, five single-scale encoders and decoders, which correspond to the five scales of semantic features output by FPN, make up the parallel structure of the multi-scale semantic encoder and decoder. Essentially, the multi-scale encoder-decoder and FPN are symbiotic structures. In this case, the encoder and decoder are not only better equipped to handle visual semantic features at different scales but also tend to preserve more valuable semantic features and productively minimize redundancy.

  After compression of the multi-scale semantic encoder, the semantic features is transmitted from the UAV to the computing server and is affected by channel noise. Additive white Gaussian noise (AWGN) channel and fading channel are considered in our work. The channel output $\hat{\boldsymbol{z}} \in \mathbb{C}^{K}$ can be expressed as

  \begin{equation}
    \label{awgn_equation}
    \hat{\boldsymbol{z}}=\boldsymbol{z}+\boldsymbol{\omega},
  \end{equation}
  where the vector $\boldsymbol{z} \in \mathbb{C}^{K}$ represents the channel input, and $K$ is the length of the vector. Vector $\boldsymbol{\omega} \in \mathbb{C}^{K}$ consists of indpendent and identically distributed samples with the distribution $\mathbb{CN}(0, \sigma^2 \mathbf{I})$ where $\sigma^2$ is the noise power and $\mathbb{CN}(\cdot,\cdot)$ is a circularly symmetric complex Gaussian distribution. The proposed system can be easily extended to other differentiable channels. In the case of a fading channel, the process can be represented as

  \begin{equation}
    \label{fading_equation}
    \hat{\boldsymbol{z}}=g\boldsymbol{z}+\boldsymbol{\omega},
  \end{equation}
  where $g \in \mathbb{C}$ denotes the channel gain. On the computing server, the multi-scale semantic decoder reconstructs the noisy multi-scale semantic features.

  Due to the inevitable noises, the semantic features expressed in the recovered messages cannot match that of the transmitter. The volume of semantic features is much less than source signals. Therefore, few transmission mistakes result in severe semantic distortion. A pioneering detection scheme is proposed to overcome these challenges. Specifically, the auxiliary knowledge of the knowledge graph and the visual semantic features are fused into a weighted graph to achieve a unified global semantic understanding of the image and improve the detection accuracy dramatically. As the cornerstone of cognitive semantic communication, the knowledge graph is a semantic network that represents relationships between entities in the form of graphs. Typically, it employs triples (head, relation, tail) or (entity, attribute, value) to convey factual information. The auxiliary knowledge can be extrapolated from factual information to enhance the performance of semantic communication systems.

\section{System Implementation}
\begin{figure*}[!t]
    \centering
    \includegraphics[width=6.8 in]{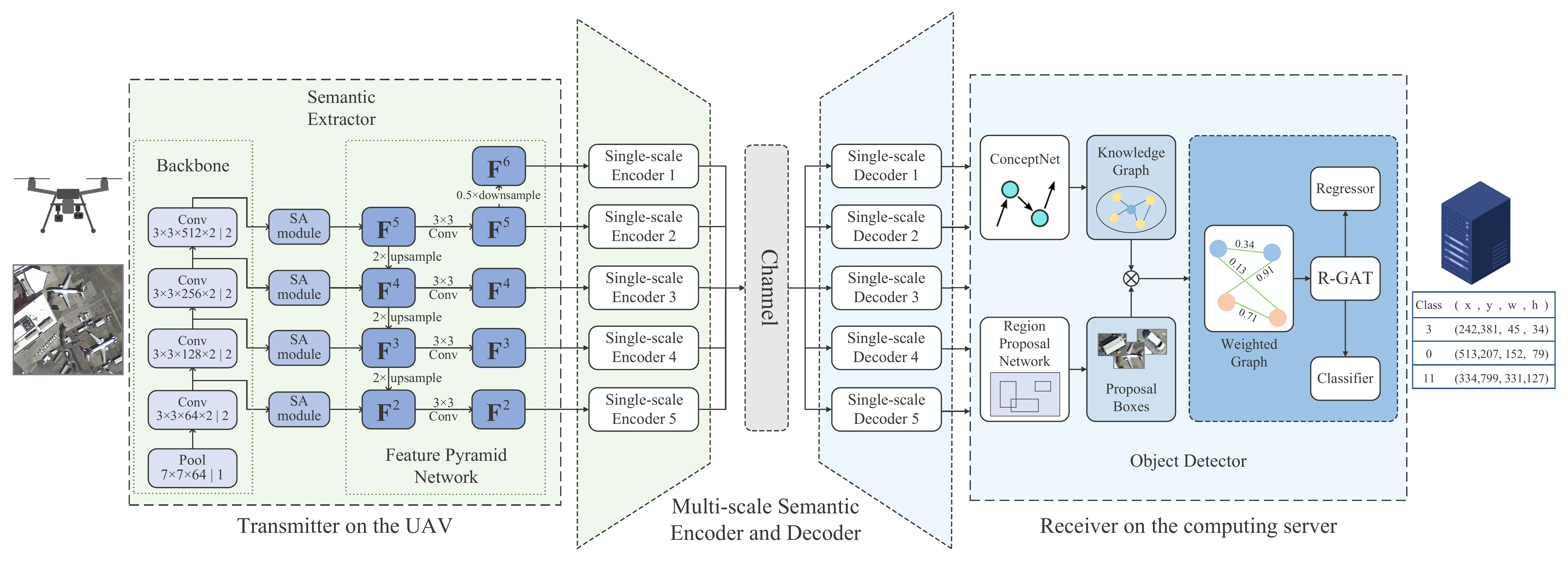}
    \caption{The network structure of the proposed UAV cognitive semantic communication system. In the backbone, Each ``Resblock'' represents a residual block. $3\times 3\times x \times 2$ indicates two convolutional layers with a kernel size of $3\times 3$ and $x$ output channels. The ``2'' following the ``\textbar'' denotes that two such convolutional structures form a single residual block. In the FPN, $F^t$ represents semantic features at different scales, $t$ denotes the index of semantic features, ``Conv'' denotes a convolutional layer, and $3\times 3$ denotes the kernel size. Note that in our system, since the image is not reconstructed at the receiver, the final output consists of the bounding box information. In the final output, ``Class'' represents the target classification category, ``x'' and ``y'' denote the coordinates of the top-left corner of the bounding box, and ``w'' and ``h'' indicate the width and height of the bounding box, respectively.}
    \label{network_structure}
  \end{figure*}

  In this section, the implementation details of the proposed system are presented. Fig. \ref{network_structure} shows the network structure of the proposed UAV cognitive semantic communication system.

  \subsection{Semantic Computing Module}
  The input of the system is the image captured by the UAV, denoted by $\mathbf{P}\in \mathbb{R}^{3\times W \times H}$, where $W$ and $H$ represent the width and height of the image, respectively. Following that, the image is inputted into the semantic extractor to extract semantic features. Considering the limited computation resources of the UAV, ResNet-18 is used as the backbone network for the semantic extractor. Note that the ResNet-18 boasts an impressive capability to extract visual semantic features because of its pre-training on ImageNet. ResNet-18 consists of four residual blocks. 
  The SA module, to be detailed in a subsequent subsection, is cascaded after each residual block to enhance the adaptability of the extracted semantic features across various SNR regimes. For aerial images, it is challenging to detect objects of different scales, especially small objects. To mitigate this issue, FPN is utilized as a semantic extractor component. 
  This approach enables the semantic extractor to obtain higher-quality multi-scale semantic features for object detection.
  Extracting semantic features can be expressed as
  
  \begin{equation}
    \mathcal{F} = \mathcal{SE}(\mathbf{P}; \alpha_T),
  \end{equation}
  where $\mathcal{F}=(\mathbf{F}^{2},\mathbf{F}^{3},\mathbf{F}^{4},\mathbf{F}^{5},\mathbf{F}^{6})$ represents the multi-scale semantic features, $\mathbf{F}^\mathcal{I}\in \mathbb{R}^{256\times S_\mathcal{I}\times S_\mathcal{I}}$. $S_\mathcal{I}$ denotes the size of semantic features. $\alpha_T$ is the trainable parameter of semantic extractor.
  
  \begin{figure}[!t]
    \centering
    \includegraphics[width=3.4 in]{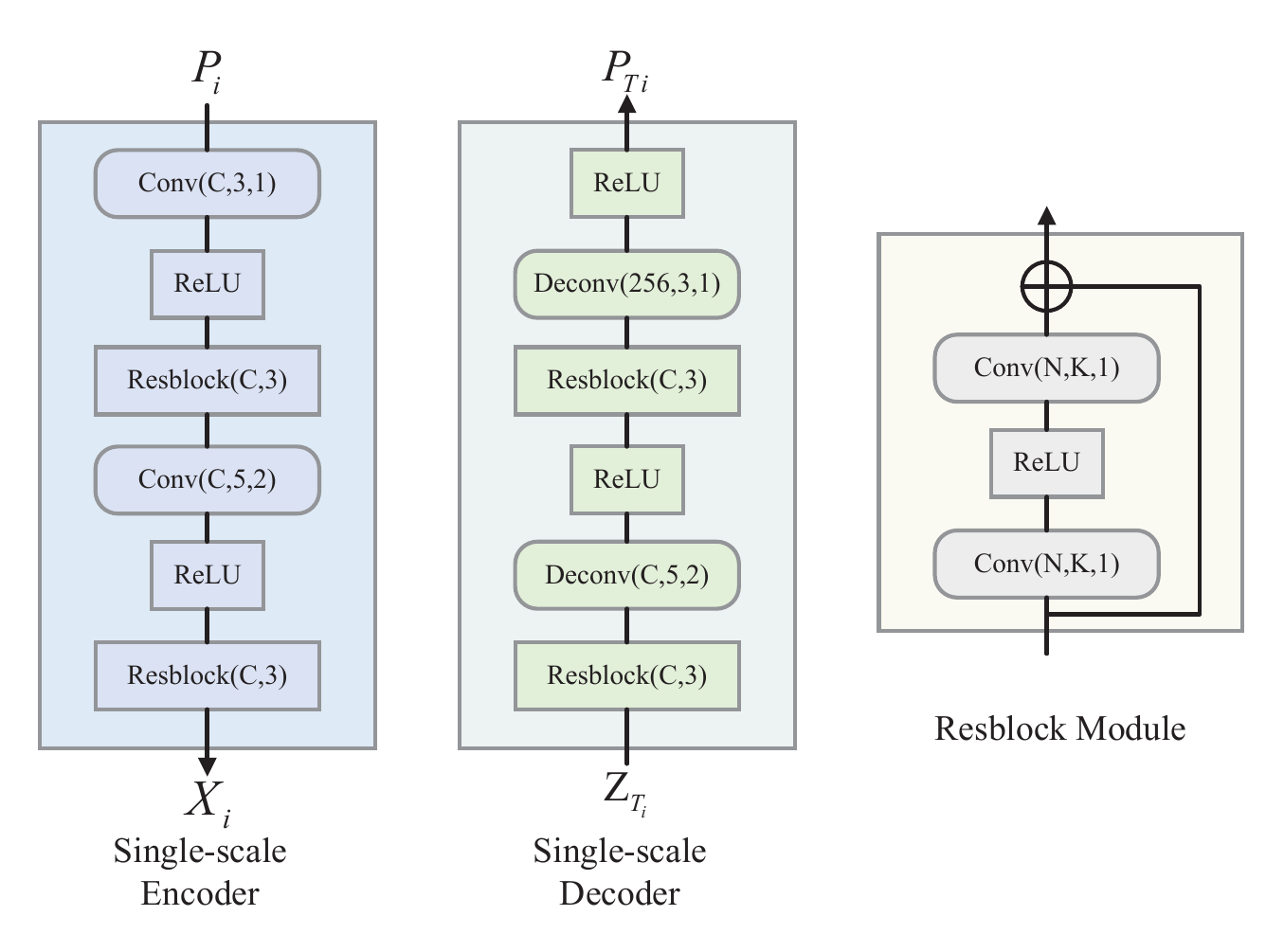}
    \caption{The details of the single-scale encoder, decoder, and residual block. “Conv(N,K,S)” and “Deconv(N,K,S)” represent the convolution and deconvolution operations with the output channel, the kernel size, and the stride as $N$, $K \times K$ and $S$, respectively.}
    \label{single_ed}
  \end{figure}

  To reduce the number of transmitted symbols, multi-scale semantic features are encoded as $\mathcal{C}=(\mathbf{C}^2, \mathbf{C}^3, \mathbf{C}^4, \mathbf{C}^5, \mathbf{C}^6)$ by using the multi-scale semantic encoder, given as

  \begin{equation}
    \mathcal{C} = \mathcal{ME}(\mathcal{F}; \beta_T),
  \end{equation}
  where $\beta_T$ is the trainable parameter of the multi-scale semantic encoder. The multi-scale semantic encoder is a parallel residual compression module that can further compress the semantic features. Each single-scale encoder is a compact auto-encoder style network, which is commonly used in computer vision for feature compression and reconstruction. The parallel design is intended to separate the transmission of multi-scale semantic features. The detailed architecture of the single-scale encoder is shown in Fig. \ref{single_ed}. In the encoder, the first convolutional layer employs a feature channel reduction strategy, while the second convolutional layer applies a scale reduction strategy, with residual blocks added to enhance compression performance. The decoder performs the reverse operations. After processing through the multi-scale semantic encoder, the dimension of semantic features is transformed as $\mathbf{C}^\mathcal{I}\in \mathbb{R}^{C\times\frac{S_\mathcal{I}}{2}\times\frac{S_\mathcal{I}}{2}}$. The number of feature channels $C$ is determined by the bandwidth compression ratio $R = k/n$, where $n$ is the size of image and $k$ is the input size of the channel.

  The encoded semantic features at each scale are reshaped as a vector and converted into the complex signal represented as $\tilde{\boldsymbol{z}} \in \mathbb{C}^{K}$. Subsequently, the vector $\tilde{\boldsymbol{z}}$ is normalized as

  \begin{equation}
    \boldsymbol{z}=\sqrt{KP}\frac{\tilde{\boldsymbol{z}}}{\sqrt{\tilde{\boldsymbol{z}}^\ast\tilde{\boldsymbol{z}}}},
  \end{equation}
  where $\tilde{\boldsymbol{z}}^\ast$ denotes the conjugate transpose of $\tilde{\boldsymbol{z}}$. The channel input $\boldsymbol{z}$ satisfies the average transmit power constraint $P$. $\boldsymbol{z}$ is transmitted over a noisy channel, which can be represented as Eq. (\ref{awgn_equation}) and Eq. (\ref{fading_equation}).

  \subsection{SNR Adaptation Module}
 
  Existing prevalent semantic communication systems are typically designed to transmit semantic features only at specific SNRs, requiring the training of several models across various SNRs. 
  Consequently, these systems must frequently switch models to maintain reliable communication in complex and dynamic environments. This model switching introduces latency, complicating the assurance of real-time communication. Additionally, the storage of multiple models requires significant memory resources.
  SNR adaptive schemes are explored within the compression-recovery semantic communication. However, they cannot be straightforwardly applied to UAV object detection tasks. The predominant reasons are the inconsistency of the system architectures and the difference in the model optimization goals. Therefore, an SA module is introduced in the proposed system to overcome the aforementioned challenges.

  \begin{figure}[t]
    \centering
    \includegraphics[width=3.5 in]{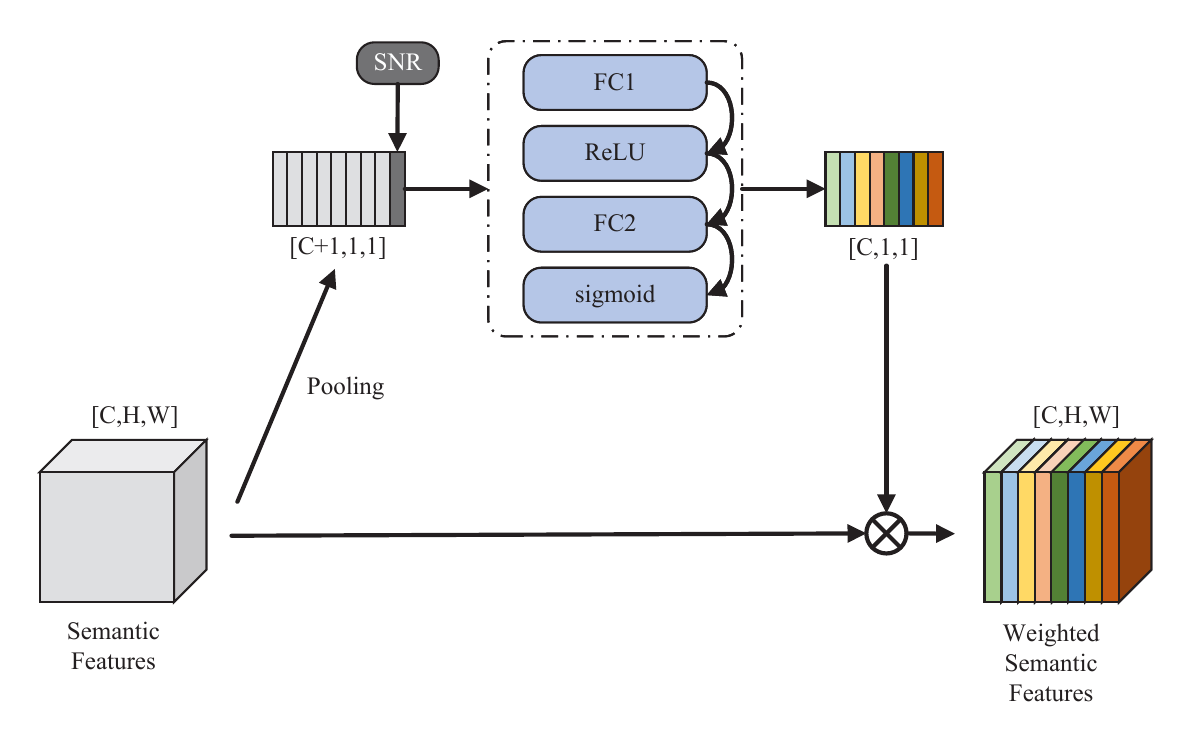}
    \caption{The details of the SA module. $C$, $H$, $W$ represent the number of channels, height, and width of the semantic features, respectively.}
    \label{snrmodule}
  \end{figure}

  The SA module is essentially a channel attention mechanism fused with contextual information \cite{hu2018squeeze}. Note that the term ``channel'' in the ``channel attention mechanism'' specifically refers to the feature channel rather than the communication channel. The attention mechanism is a deep learning technique widely used in natural language processing and computer vision. These mechanisms employ an additional neural network to selectively identify certain features or to assign varying weights to different features.
  In our work, the channel attention mechanism evaluates the importance of each channel in the feature map and assigns weights to it. Therefore, the more effective feature channels for the task are amplified, while those with poor contribution are suppressed. When calculating attention weights, the SA module introduces SNR as context information to enhance the adaptability of semantic features to noise interference under different SNRs. 
  
  The SA module is integrated into the backbone network, ResNet-18, which has four basic blocks corresponding to multi-scale semantic features. After each basic block, an SA module is deployed to assign attention weights to the corresponding scale of semantic features. Since this design is implemented in the semantic extraction network at the transmitter, it has high scalability for performing various visual tasks in task-oriented semantic communication.

  The detailed architecture of the SA module is depicted in Fig. \ref{snrmodule} and Algorithm \ref{alg_samodule}. The output of the $i$th basic block is semantic features  $\mathbf{O}^\mathcal{I}\in \mathbb{R}^{256\times S_\mathcal{I}\times S_\mathcal{I}}$, comprising several feature channels. The semantic features $\mathbf{O}^\mathcal{I}$ undergo processing and conversion into a one-dimensional vector $\boldsymbol{p}$ through a global average pooling function. After that, the one-dimensional vector $\boldsymbol{p}$ is concatenated with SNR to form the context information $\boldsymbol{p}^\mu$. The context information is fed into several fully connected layers to produce the weight factor $\boldsymbol{s}$. The weighted features $\mathbf{Q}^\mathcal{I}$ are calculated by multiplying the features $\mathbf{O}^\mathcal{I}$ and the weight factor $\boldsymbol{s}$. This outcome in different scaled features $\mathbf{Q}^\mathcal{I}$ depends on the exact SNR regime. Finally, $\mathbf{Q}^\mathcal{I}$ are fed into the FPN and the single-scale features $\mathbf{F}^\mathcal{I}$ are derived.

  \subsection{Knowledge Graph Enhanced Object Detection}
  At the computing server, the channel output $\hat{\boldsymbol{z}}$ is converted to the real number and moulded to the original dimension like $\mathcal{C}$ before reconstructing the semantic features. The transformed semantic features is denoted as $\hat{\mathcal{C}}$. Following that, the multi-scale semantic decoder executes the reconstruction procedure, given as
  
  \begin{equation}
    \hat{\mathcal{F}} = \mathcal{MD}(\hat{\mathcal{C}}; \beta_R),
  \end{equation}
  where $\beta_R$ is the trainable parameter of multi-scale semantic decoder. The multi-scale semantic decoder shares a symmetrical structure with the multi-scale semantic encoder and uses deconvolutional layers to reconstruct the semantic features. The details of the single-scale decoder are demonstrated in Fig. \ref{single_ed}.

  \begin{algorithm}[!t]
    \KwIn{The semantic features $\mathbf{O}^\mathcal{I}$, the SNR information $\mu$, the weight of the fully connected layers $\mathbf{W}_1$ and $\mathbf{W}_2$, the bias of the fully connected layers $b_1$ and $b_2$.}
    \KwOut{The weighted semantic features  $\mathbf{Q}^\mathcal{I}$.}

    $c$ = size$(\mathbf{O}^\mathcal{I})$ \\
    \tcp{Calculate the number of feature channels in the semantic features}
    $\boldsymbol{p}$ = GlobalAveragePooling$(\mathbf{O}^\mathcal{I})$ \\
    \tcp{Perform the average global pooling function}
    $\boldsymbol{p}^{\mu}$ = concatenate$(\boldsymbol{p}, \mu)$ \\
    \tcp{Calculate the context information}
    $\boldsymbol{s}$ = sigmoid$(\mathbf{W}_2\cdot $ReLU$(\mathbf{W}_1\boldsymbol{p}^{\mu}+b_1)+b_2)$ \\
    \tcp{Calculate the weight factor}
    Convert the weight factor $\boldsymbol{s}$ to the channel-wise weight factor $\boldsymbol{s}_i$, $i=0, 1, 2, ..., c-1$
    
    Convert the semantic features $\mathbf{O}^\mathcal{I}$ to the channel-wise features $\mathbf{O}^\mathcal{I}_{i,:,:}, i = 0, 1, 2, ..., c-1$

    \For{$k \leftarrow 0$ \KwTo $c-1$} {
      $\mathbf{Q}^\mathcal{I}_{i,:,:} = \boldsymbol{s}_i \cdot  \mathbf{O}^\mathcal{I}_{i,:,:}$ \\
      \tcp{Calculate the semantic channel-wise weighted features}
    }
    Convert the channel-wise weighted semantic features $\mathbf{Q}^\mathcal{I}_{i,:,:}$, $i=0,1,2,...,c-1$, to the weighted semantic features $\mathbf{Q}^\mathcal{I}$
    \caption{SA module.}\label{alg_samodule}
  \end{algorithm}

  \begin{figure}[!t]
    \centering
    \includegraphics[width=3.4 in]{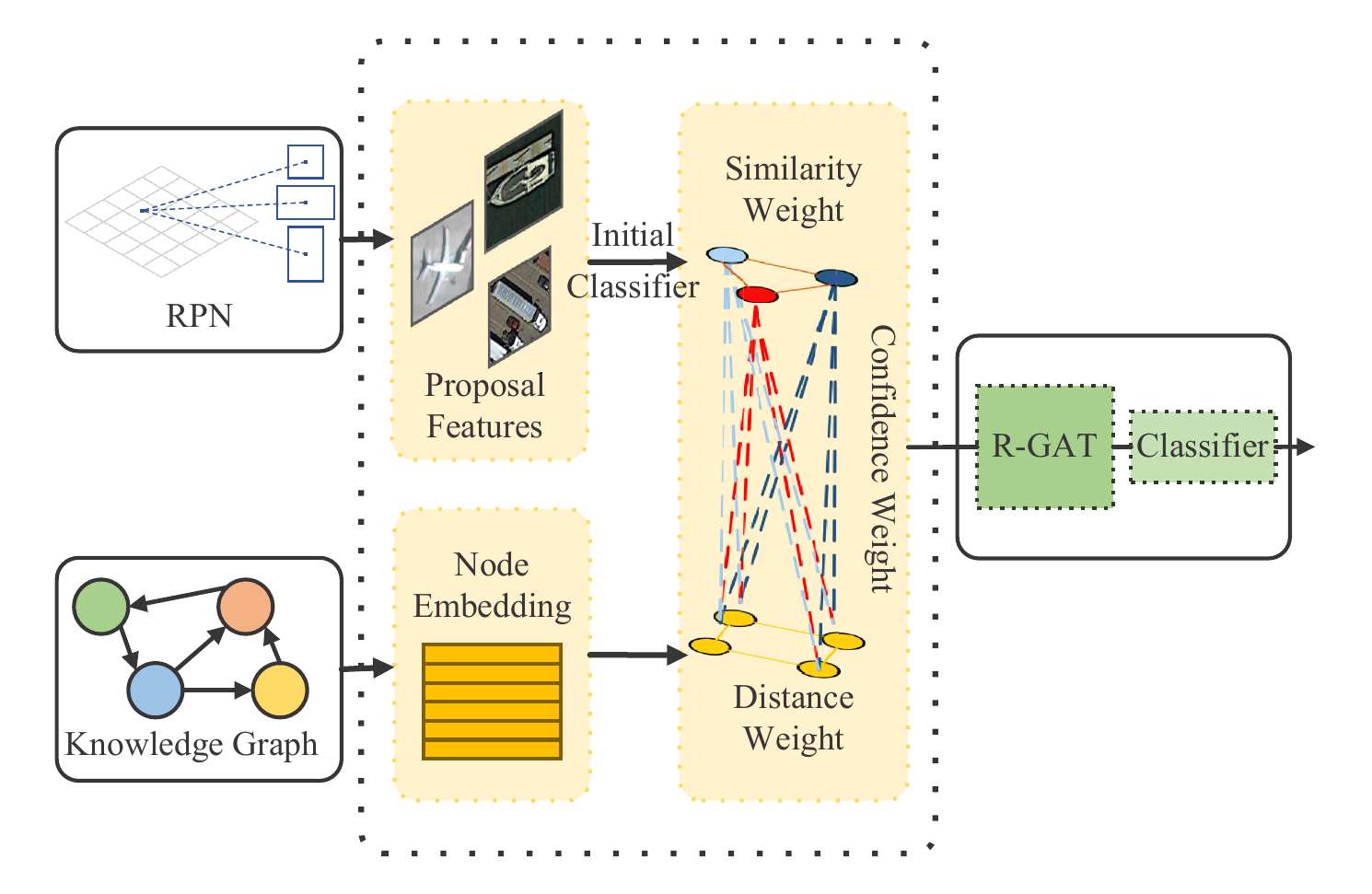}
    \caption{The scheme of exploiting the auxiliary knowledge of the knowledge graph to improve the system performance.}
    \label{detection_scheme}
  \end{figure}

  \begin{figure*}[!t]
    \centering
    \includegraphics[width=5.7 in]{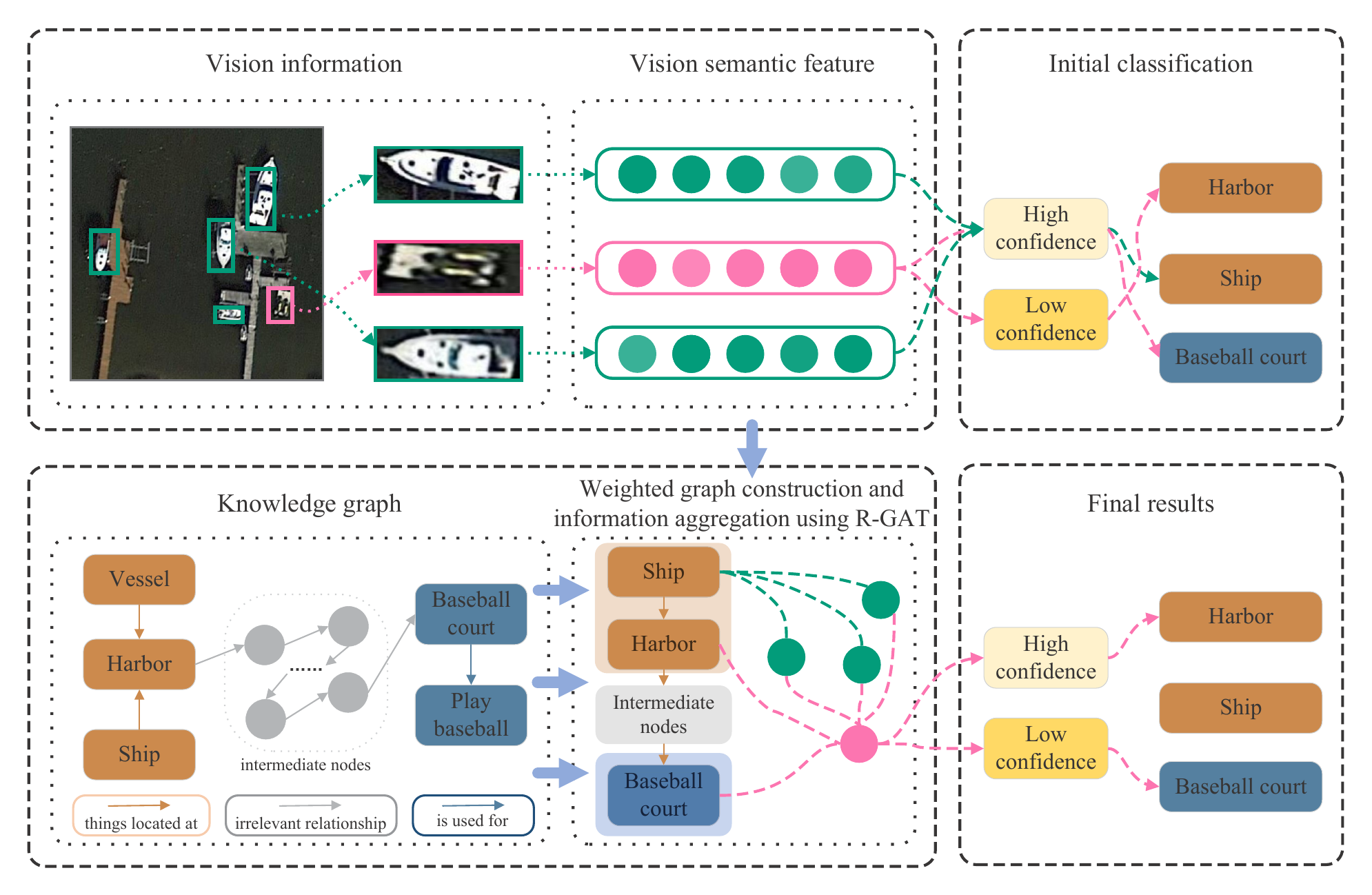}
    \caption{Example of the detection scheme applied to an image. Note that in the weighted graph, pink nodes and green nodes represent visual semantic features. In practice, all nodes in the weighted graph exist as vector features. For visual clarity, nodes from the knowledge graph are presented in natural language form.}
    \label{example}
\end{figure*}

  The object detector is the core component of the proposed system, utilizing the noisy semantic features it receives to achieve the final detection outcomes. Compared to conventional object detection, semantic communication-driven object detection systems usually exhibit slightly lower detection accuracy. It is attributed to compression distortion and noise interference, which can lead to misjudgments of objects by the object detector. Therefore, we provide a novel object detection scheme driven by a knowledge graph, which significantly alleviates this issue. Unlike existing knowledge graph-based object detection research, a novel method is proposed of constructing a weighted graph to incorporate knowledge graph information. The weighted graph includes knowledge graph node embeddings, visual semantics, node embedding distances, visual semantic similarities, and initial classification confidence information. A relational graph attention network aggregates the semantic information in the weighted graph, producing high-accuracy results through training.

  The knowledge graph contains abundant auxiliary knowledge and has the capability to establish relationships and similarities among various objects in the image. To construct our specialized knowledge graph, a subgraph extraction approach is utilized from a larger, pre-existing knowledge graph, named ConceptNet. ConceptNet is a freely available semantic network consisting of numerous concepts, descriptions, and common-sense relationships. The introduction of extensive external information while building the knowledge graph can lead to redundancy, which may affect the performance of the system. Therefore, data cleaning and knowledge fusion are employed for quality control during the construction. Specifically, our construction method consists of three stages. In the first stage, a subgraph specifically pertaining to the relevant entities of detection categories is extracted. To achieve this, key entities and relationships are identified within detection categories. Then, graph querying techniques are applied to extract our knowledge graph from ConceptNet, ensuring that only relevant entities and their connections are included. Data cleaning in the second stage is a crucial step to maintain the integrity and quality of our knowledge graph. The removal of isolated nodes, low-degree nodes, and duplicate nodes is involved in this stage. In the third stage, knowledge fusion techniques are employed to reduce redundancy at the semantic level of our knowledge graph. The resulting knowledge graph consists of 3,579 unique entities connected by 31 types of relationships, providing a detailed and focused knowledge base for our system.

  To further fuse the semanic information and enhance the global semantic understanding of system, a weighted graph is constructed. The weighted graph encompasses two categories of nodes and three types of weighted edges, with the goal of exploiting global semantic information. Note that the original information is preserved in the weighted graph. Therefore, the accuracy of the results is only enhanced compared to that without introducing additional information.

  \begin{algorithm}[!h]

    \KwIn{The reconstructed semantic features $\hat{\mathcal{F}}$, the node embeddings of knowledge graph $\mathbf{E}$, the region proposal network RPN$(\cdot)$, the initial classifier $fc_{init\_cls}(\cdot)$, the relational graph attention network RGAT$(\cdot)$, the final classifier $fc_{cls}(\cdot)$ and regressor $fc_{reg}(\cdot)$.}
    \KwOut{The detection result $\mathbf{D}^{cls}$ and $\mathbf{D}^{reg}$.}

    $\mathbf{B}$ = RPN$(\hat{\mathcal{F}})$ \\
    \tcp{Generate the semantic features of proposal boxes from the reconstructed semantic features}

    $\mathbf{M}^\mathcal{C}$ = $fc_{init\_cls}(\mathbf{B})$ \\
    \tcp{Calculate the prediction matrix of the initial classification}

    \For{$i \leftarrow 0$ \KwTo $M-1$} {
      \For{$j \leftarrow 0$ \KwTo $K-1$} {
        \If{$\mathbf{M}^\mathcal{C}_{i,j} \in $ top3$(\mathbf{M}^\mathcal{C}_{i,:})$} {
          \tcp{Link the top-3 predictions with the knowledge graph nodes}
          $link(\mathbf{B}_{i,:}, \mathbf{E}_{j,:}, \mathbf{M}^\mathcal{C}_{i,j})$ \\
          \tcp{Link the proposal boxes and the knowledge graph nodes}
        }
      }
    }

    \For{$i \leftarrow 0$ \KwTo $M-1$} {
      \For{$j \leftarrow 0$ \KwTo $M-1$} {
        \If{$i \neq j$} {
          $link(\mathbf{B}_{i,:}, \mathbf{B}_{j,:}, D_v(\mathbf{B}_{i,:}, \mathbf{B}_{j,:}))$ \\
          \tcp{Link the proposal boxes}
        }
      }
    }

    \For{$i \leftarrow 0$ \KwTo $K-1$} {
      \For{$j \leftarrow 0$ \KwTo $K-1$} {
        \If{$i \neq j$} {
          $link(\mathbf{E}_{i,:}, \mathbf{E}_{j,:}, D_v(\mathbf{E}_{i,:}, \mathbf{E}_{j,:}))$ \\
          \tcp{Link the knowledge graph nodes}
        }
      }
    }

    Organize the linked information into a graph $\mathcal{G}$

    $\mathbf{B}^\mathcal{E}$ = RGAT$(\mathcal{G})$ \\
    \tcp{Calculate the enhanced semantic features of proposal boxes}

    $\mathbf{D}^{cls}, \mathbf{D}^{reg} = fc_{cls}(\mathbf{B}^\mathcal{E}), fc_{reg}(\mathbf{B}^\mathcal{E})$ \\
    \tcp{Calculate the detection result}

    \caption{Knowledge Graph Enhanced Object Detection.}\label{alg_kg}
  \end{algorithm}

  The proposed detection scheme is shown in Fig. \ref{detection_scheme}, and the detailed algorithm is outlined in Algorithm \ref{alg_kg}. The object detector is supplied with the reconstructed semantic features $\hat{\mathcal{F}}$ to obtain the results. A region proposal network (RPN) and a knowledge graph comprise the object detector. The RPN is in a position to generate proposal boxes $\mathbf{B}\in \mathbb{R}^{M\times N}$,  $M$ is the number of proposal boxes, and $N$ is the dimension of the features. Following that, the proposal boxes are fed into an initial classifier to obtain the initial prediction matrix $\mathbf{M}^\mathcal{C}\in\mathbb{R}^{M\times C}$, $C$ is the number of categories, which is utilized to establish weights with nodes in the knowledge graph. Simultaneously, the weights between proposal boxes are derived from the similarity function. The knowledge graph is responsible for introducing auxiliary knowledge to build relationships between objects in proposal boxes. To extract the auxiliary knowledge of the knowledge graph, Metapath2vec \cite{guo2022asurvey} model is adopted to embed the nodes of the knowledge graph. Metapath2vec is a model that learns structural and semantic knowledge from a large heterogeneous graph in an unsupervised manner, and it has been widely used for graph representation learning. Since Metapath2vec can learn embedding representations that map nodes to a low-dimensional embedding space, nodes with similar structural roles or semantic relationships are clustered together. The node embeddings are denoted as $\mathbf{E}\in \mathbb{R}^{K\times N}$, $K$ is the number of nodes, and $N$ is the dimension of node embedding. Subsequently, the weights between nodes in the knowledge graph are derived through an embedding distance function. The embedding distance function and the similarity function are calculated through the cosine similarity function, which can be represented by 

  \begin{equation}
      D_v(\boldsymbol{u}, \boldsymbol{v}) = \frac{\boldsymbol{u} \cdot \boldsymbol{v}}{\lVert \boldsymbol{u} \rVert \lVert \boldsymbol{v} \rVert},
  \end{equation}
  where $\boldsymbol{u}$ and $\boldsymbol{v}$ represent the embeddings of distinct nodes.

  After constructing the weighted graph, a relational graph attention network (R-GAT) \cite{wang2020relational} is applid to unify the global semantic information of an image. R-GAT extend the graph attention network model to handle relational data, where multiple types of edges or relationships exist between nodes in a graph. R-GAT enhances node representations by considering not only the features of neighboring nodes but also the types of relations between them, applying attention mechanisms specific to each relation type. This allows the model to focus on the most relevant neighbors and relations when updating node embeddings. In our work, three types of relational edges are defined, namely, from knowledge graph nodes to other knowledge graph nodes, from knowledge graph nodes to proposal boxes, and from proposal boxes to other proposal boxes. We map the edge weights to the same dimensionality as the node embeddings using a simple, fully connected network. Through R-GAT, the knowledge graph node embeddings and the semantic features of the proposal boxes are both enhanced. The enhanced proposal boxes are subsequently placed into a classifier to provide final results. As illustrated in Fig. \ref{example}, it is challenging to differentiate the object within the pink bounding box based on the semantic features of the object in the initial classification, leading to misjudgment. By exploiting classification results of other objects in the same image and auxiliary knowledge of the knowledge graph, the system establishes a comprehensive global semantic understanding of the image, thereby enabling correct classification.

  \subsection{Loss Function}
  To optimize the parameters of the proposed system, we minimize the multi-task loss function, which is defined as
  \begin{align}
      L(\{p_i\}, \{t_i\})=\frac{1}{N_{cls}}\sum_{i}L_{cls}(p_i, p_i^\ast) \notag
      \\+ \lambda\frac{1}{N_{reg}}\sum_{i}p_i^\ast L_{reg}(t_i, t_i^\ast), 
  \end{align}
  where $i$ is the index of an anchor in a mini-batch and $p_i$ is the predicted probability of anchor $i$ being an object. The ground truth label $p_i^\ast$ is 1 if the anchor is positive, and is 0 if the anchor is negative. $t_i$ is a vector representing the 4 parameterized coordinates of the predicted bounding box, and $t_i^\ast$ is that of the ground-truth box associated with a positive anchor. The classification loss $L_{cls}$ is log loss. The regression loss $L_{reg}(t_i, t_i^\ast)=R(t_i-t_i^\ast)$, where $R$ is a smooth $L_1$ function. Using this loss function, all learnable parameters of the system are optimized.

\section{Simulation Results}
  In this section, the simulation is conducted to evaluate the performance and efficacy of the proposed system across various communication channels, bandwidth compression ratios, and SNRs. Additionally, an ablation study is undertaken to characterize the effectiveness of the proposed methods.

  \subsection{Dataset}

  \begin{table}[h]
    
    \renewcommand{\arraystretch}{1.5}
    \centering
    \caption{The number of samples in the training and testing sets.}
    \setlength{\tabcolsep}{1 cm}
    \begin{tabular}{cc}
    \hline
    \textbf{Dataset} & \textbf{Number of Samples} \\
    \hline
    Training set & 15750  \\
    Testing set & 5298 \\
    \hline
    \end{tabular}
    \label{dataset_split}
  \end{table}

  The used dataset is the public large-scale dataset for object detection in aerial images (DOTA) \cite{zhu2022detection}, which is a well-known benchmark object detection dataset in aerial scenes comprising 2806 images. Each image is approximately $4000 \times 4000$ pixels in size and contains objects exhibiting a wide variety of scales. The dataset is annotated with a total of 188,282 instances across 15 categories, including \emph{small vehicle}, \emph{large vehicle}, \emph{plane}, \emph{storage tank}, \emph{ship}, \emph{harbor}, \emph{ground track field}, \emph{soccer ball field}, \emph{tennis court}, \emph{swimming pool}, \emph{baseball diamond}, \emph{roundabout}, \emph{basketball court}, \emph{bridge}, and \emph{helicopter}. We preprocessed the dataset by segmenting large-sized images into $1024 \times 1024$ pixels. Moreover, to mitigate potential confounding factors and ensure consistent system performance, the oriented bounding boxes of the original dataset are transformed into horizontal bounding boxes. Subsequently, the dataset is split into training and testing sets. The number of samples in the training and testing set is presented in Table \ref{dataset_split}.

  \subsection{Benchmarks and Simulation Setup}

  \begin{table}[h]
    \renewcommand{\arraystretch}{1.5}
    \centering
    \caption{The simulation parameter setting.}
    \setlength{\tabcolsep}{0.7 cm}
    \begin{tabular}{cc}
    \hline
    \textbf{Parameters} & \textbf{Values} \\
    \hline
    Batch size & 16 \\
    Momentum & 0.9 \\
    Weight decay & $1\times 10^{-4}$ \\
    Initial learning rate & $1\times 10^{-2}$ \\
    Learning rate decay &  0.3 \\
    Loss function & Multi-task loss \\
    Optimizer & Adam \\
    Number of epochs & 70 \\
    NMS threshold & 0.5 \\
    Maximum detection number (training) & 1024 \\
    Maximum detection number (testing) & 100 \\
    \hline
    \end{tabular}
    \label{parameter_setting}
  \end{table}

  For performance comparison, we provide the following two benchmarks. The communication method for the two benchmarks vary, and the object detection algorithm both utilize Faster R-CNN with ResNet-18 for a consistent comparison.

  \begin{figure*}[t]
    \centering
    \subfloat[AWGN channel.]{
      \includegraphics[width=3.3 in]{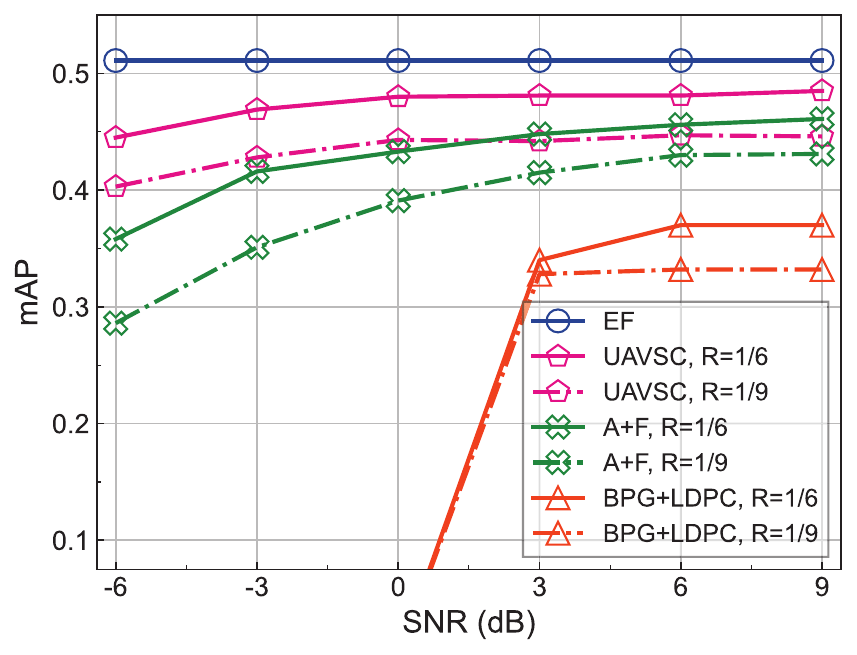}
      \label{cr6_9_awgn}}
    \subfloat[Rayleigh channel.]{
      \includegraphics[width=3.3 in]{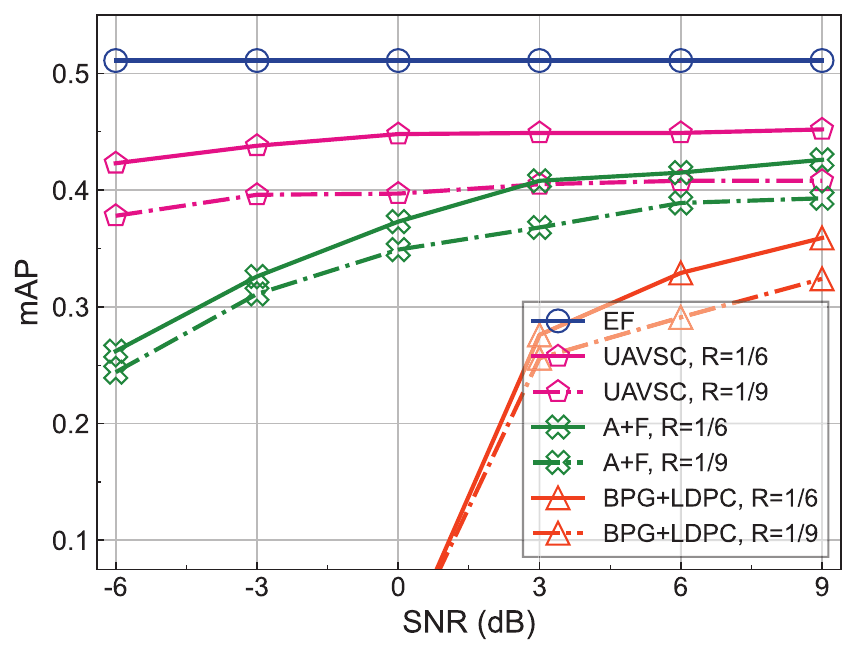}
      \label{cr6_9_rayleigh}}
    \caption{mAP versus SNR for the proposed system and two benchmarks under AWGN channel and Rayleigh channel.}
    \label{cr6_9map}
  \end{figure*}
  
  \emph{1) Benchmark 1:} The first benchmark is a conventional communication system. We employ the better portable graphics (BPG) format for source coding and LDPC for channel coding. BPG format is an image format based on HEVC video codec and is one of the most efficient image compression methods among the non-neural network-based image codes. After multiple rounds of evaluation in terms of the best performance, the selected parameters for coding rate and modulation scheme are determined to be 1/3 and 16-QAM, respectively.

  \emph{2) Benchmark 2:} The second benchmark is a deep learning-powered semantic communication system, ADJSCC. In this system, for improved transmission and recovery of large-sized aerial images, we increase the number of convolutional filters in both the encoder and decoder. Additionally, the filter size is modified from 5 to 7. Other hyperparameters follow the same settings in \cite{xu2021wireless} unless otherwise specified. 

  Among all simulation results, EF represents error-free transmission, which means the object detection is directly performed through Faster R-CNN with ResNet-18. Therefore, the precision is constant and not affected by channel conditions. Moreover, we adopt the ADJSCC+Faster R-CNN denoted as A+F and the conventional communication system denoted as BPG+LDPC. UAVSC is the abbreviation of our proposed system. 

  To efficiently extract semantic features while conserving computation resources at the transmitter, a ResNet-18 with FPN pre-trained from the MS COCO dataset is adopted as the base model in our proposed system. SA modules are subsequently incorporated after each residual block. The number of convolutional neural network filters of the semantic encoder correlates with the bandwidth compression ratio. For the detector at the receiver, a three-layer R-GAT model with two-head attention is employed, where the dimensions of the input and output layers are both set to 1024, and the hidden layer has a dimension of 512. During the training stage, the learning rate is set to 0.01, the batch size is set to 16, and the epoch is configured as 70. With the increase in training epoch, the learning rate undergoes three decay steps at a rate of 0.3 to promote the convergence of the system. The Adam optimizer is used to train the entire system. In the testing stage, the non-maximum suppression (NMS) threshold and max detection number are set to 0.5 and 100, respectively. The specific simulation parameter setting is shown in Table \ref{parameter_setting}.

  For consistent performance comparisons, all systems run on identical hardware. The simulated environment consists of a Linux server equipped with two AMD EPYC 7302 16-core CPUs and an NVIDIA GeForce RTX 3090 GPU.

  \subsection{Performance Metric}
  The primary metrics in subsequent analyses align with the standards set by the MS COCO dataset, with a particular emphasis on utilizing average precision (AP) at an intersection-over-union (IoU) threshold of 0.5, namely AP$_{50}$. It is essential to clarify that in the context of our work, the term mAP holds a distinct meaning from some object detection literature. Specifically, mAP refers to the mean values of AP$_{50}$ rather than AP$_{50:95}$ across all categories. AP$_S$, AP$_M$, and AP$_L$ also appear as evaluation metrics, indicating the average precision for small-scale, medium-scale and large-scale objects, respectively. Precisely, the pixel size of the small-scale objects is defined as smaller than $32\times 32$, the pixel size of the medium-scale objects falls between $32\times 32$ and $96\times 96$, and the pixel size of the large-scale objects is larger than $96\times 96$. The IoU threshold is also 0.5 when calculating AP$_S$, AP$_M$, and AP$_L$.

  \subsection{Performance Comparison}

  \begin{table*}[b]
    \renewcommand{\arraystretch}{1.5}
    \centering
    \setlength{\tabcolsep}{0.03 cm}
    \caption{The detailed evaluation of systems in specific conditions.}
    \begin{tabular}{cccccccc}
    \hline
    Condition & Method & AP$_{50:95}$ & AP$_{50}$ & AP$_{75}$ & AP$_S$ & AP$_M$ & AP$_L$\\
    \hline
    \multirow{2}{*}{AWGN channel, R $=$ 1/6, SNR $=$ -6dB} & UAVSC & $\textbf{0.252}(+29.2\%)$ & $\textbf{0.445}(+24.3\%)$ & $\textbf{0.242}(+10.5\%)$ & $\textbf{0.147}(+33.6\%)$ & $\textbf{0.376}(+23.6\%)$ & $\textbf{0.626}(+46.6\%)$\\
                           & A+F & $0.195$ & $0.358$ & $0.219$ & $0.110$ & $0.304$ & $0.427$\\
    \hline
    \multirow{2}{*}{AWGN channel, R $=$ 1/6, SNR $=$ 0dB} & UAVSC & $\textbf{0.280}(+11.6\%)$ & $\textbf{0.480}(+10.9\%)$ & $\textbf{0.282}(+11.9\%)$ & $\textbf{0.174}(+23.4\%)$ & $\textbf{0.408}(+2\%)$ & $\textbf{0.636}(+9.8\%)$\\
                                  & A+F & $0.251$ & $0.433$ & $0.252$ & $0.141$ & $0.400$ & $0.579$\\
    \hline
    \multirow{3}{*}{AWGN channel, R $=$ 1/6, SNR $=$ 9dB} & UAVSC & $\textbf{0.284}(+2.5\%)$ & $\textbf{0.485}(+5.2\%)$ & $\textbf{0.288}(+2.9\%)$ & $\textbf{0.181}(+10.4\%)$ & $0.413(-4.8\%)$ & $\textbf{0.651}(+2.2\%)$\\
                                  & A+F & $0.277$ & $0.461$ & $0.280$ & $0.164$ & $\textbf{0.434}$ & $0.637$\\
                                  & BPG+LDPC & $0.194$ & $0.370$ & $0.193$ & $0.099$ & $0.303$ & $0.477$\\
    \hline
    \multirow{2}{*}{Rayleigh channel, R $=$ 1/6, SNR $=$ -6dB} & UAVSC & $\textbf{0.235}(+51.6\%)$ & $\textbf{0.423}(+61.5\%)$ & $\textbf{0.227}(+38.4\%)$ & $\textbf{0.130}(+88.4\%)$ & $\textbf{0.380}(+63.8\%)$ & $\textbf{0.583}(+61\%)$\\
                           & A+F & $0.155$ & $0.262$ & $0.164$ & $0.069$ & $0.232$ & $0.362$\\
    \hline
    \multirow{2}{*}{Rayleigh channel, R $=$ 1/6, SNR $=$ 0dB} & UAVSC & $\textbf{0.252}(+13.5\%)$ & $\textbf{0.448}(+20.1\%)$ & $\textbf{0.249}(+11.2\%)$ & $\textbf{0.143}(+18.2\%)$ & $\textbf{0.391}(+18.1\%)$ & $\textbf{0.597}(+11.2\%)$\\
                                  & A+F & $0.222$ & $0.373$ & $0.224$ & $0.121$ & $0.331$ & $0.537$\\
    \hline
    \multirow{3}{*}{Rayleigh channel, R $=$ 1/6, SNR $=$ 9dB} & UAVSC & $\textbf{0.257}(+2\%)$ & $\textbf{0.452}(+6.1\%)$ & $0.250(-2.7\%)$ & $\textbf{0.148}(+2.1\%)$ & $0.399(-1\%)$ & $0.591(-0.7\%)$\\
                                  & A+F & $0.252$ & $0.426$ & $\textbf{0.257}$ & $0.145$ & $\textbf{0.403}$ & $\textbf{0.595}$\\
                                  & BPG+LDPC & $0.187$ & $0.356$ & $0.204$ & $0.095$ & $0.299$ & $0.405$\\
    \hline
    \end{tabular}
    \label{apcomprision}
  \end{table*}

  Fig. \ref{cr6_9map} displays mAP versus SNR for the proposed system and two benchmarks under AWGN channel and Rayleigh channel, while R $=$ 1/6 and 1/9. According to Fig. \ref{cr6_9map}, an evident conclusion can be drawn that the proposed system outperforms the two benchmarks under the same channel conditions, especially at low SNRs. Specifically, the proposed system achieves an average improvement of 4.48\% and 7.48\% over the A+F system at SNR $=$ -6dB and 5.08\% and 5.63\% at SNR $=$ 9dB in AWGN and Rayleigh channels, respectively. Moreover, the stability of the proposed system at different SNRs indicates its robustness to inevitable noises. 
  This robustness can be attributed, in part, to the inherent benefits of task-oriented semantic transmission. In addition, the integration of the knowledge graph and the SA module also contribute significantly to the detection performance.
  Note that the conventional communication system results in data transmission failure when SNR $\leq$ 0dB under AWGN and Rayleigh channels, leading to mAP being 0. Even at relatively higher SNRs, the performance of the conventional system remains unsatisfactory because of the severe image distortion caused by the source coding stage. Fig. \ref{cr6_9_awgn} and Fig. \ref{cr6_9_rayleigh} showcase the performance of systems in AWGN and Rayleigh channels, respectively. While SNR $\leq$ 0dB, the mAP of the proposed system at R $=$ 1/9 surpasses that of A+F at R $=$ 1/6, highlighting the ability of the proposed system to transmit less data while maintaining precision. The superiority of the proposed system at low bandwidth compression ratios is due to the multi-scale semantic encoder and decoder, which alleviates compression distortion through parameter learning.

  \begin{figure*}[h]
    \centering
    \subfloat[AWGN channel.]{
      \includegraphics[width=3.3 in]{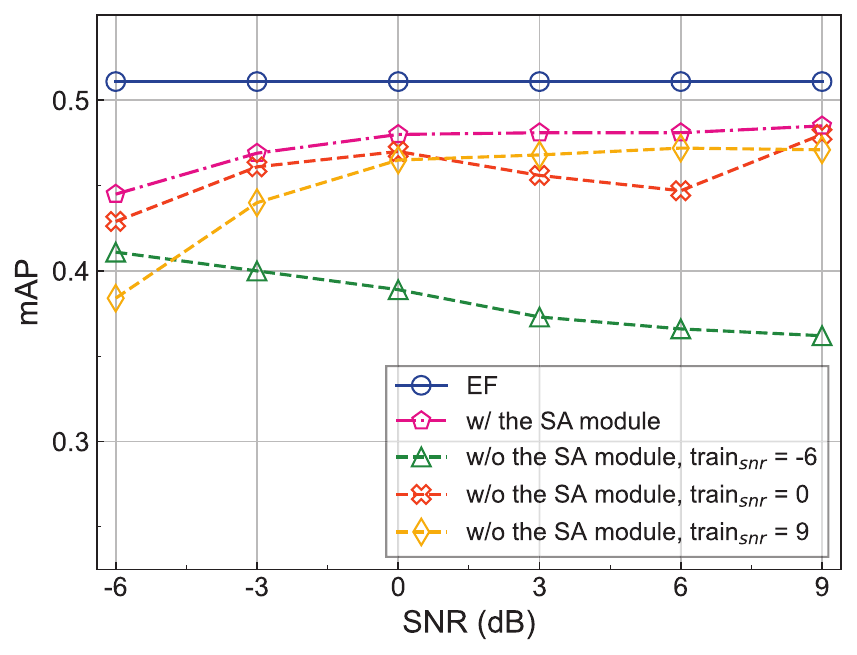}
      \label{cr6_awgn_SAmodule_ablation}}
    \subfloat[Rayleigh channel.]{
      \includegraphics[width=3.3 in]{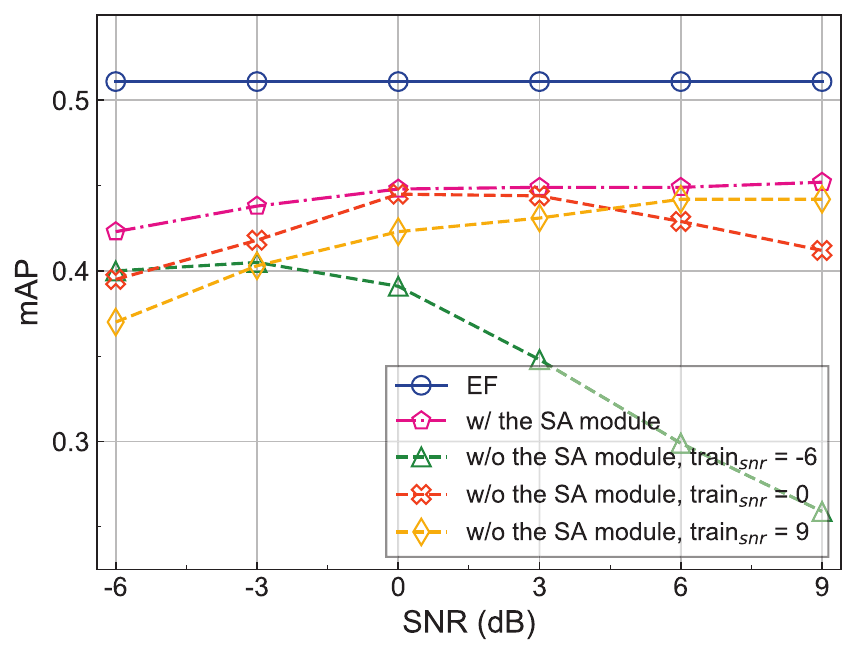}
      \label{cr6_rayleigh_SAmodule_ablation}}
    \caption{mAP versus SNR for the proposed system with and without the SA module under AWGN and Rayleigh channels.}
    \label{cr6_SAmodule_ablation}
  \end{figure*}

  \begin{figure*}[t]
    \centering
    \subfloat[AWGN channel.]{
      \includegraphics[width=3.3 in]{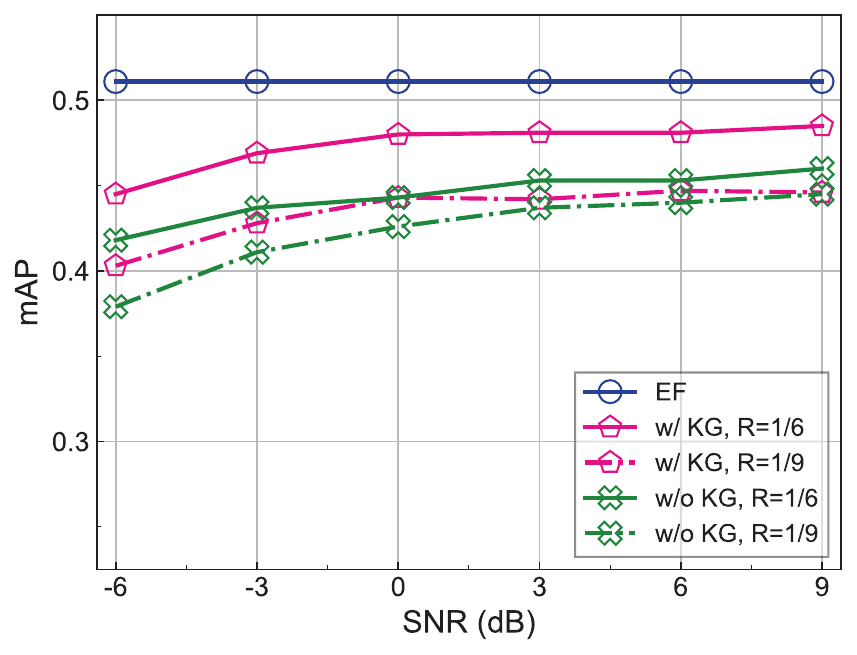}
      \label{cr6_awgn_kg_ablation}}
    \hspace{0.25in}
    \subfloat[Rayleigh channel.]{
      \includegraphics[width=3.3 in]{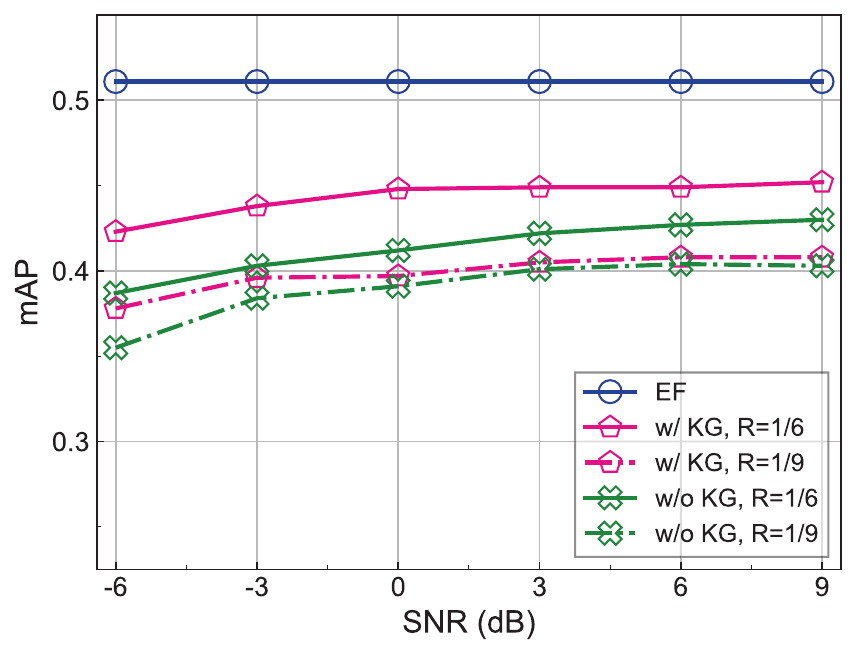}
      \label{cr6_rayleigh_kg_ablation}}
    \caption{mAP versus SNR for proposed system with and without the KG under AWGN and Rayleigh channels.}
    \label{cr6_kg_ablation}
  \end{figure*}

  \begin{figure*}[t]
    \centering
    \subfloat[AWGN channel.]{
      \includegraphics[width=3.3 in]{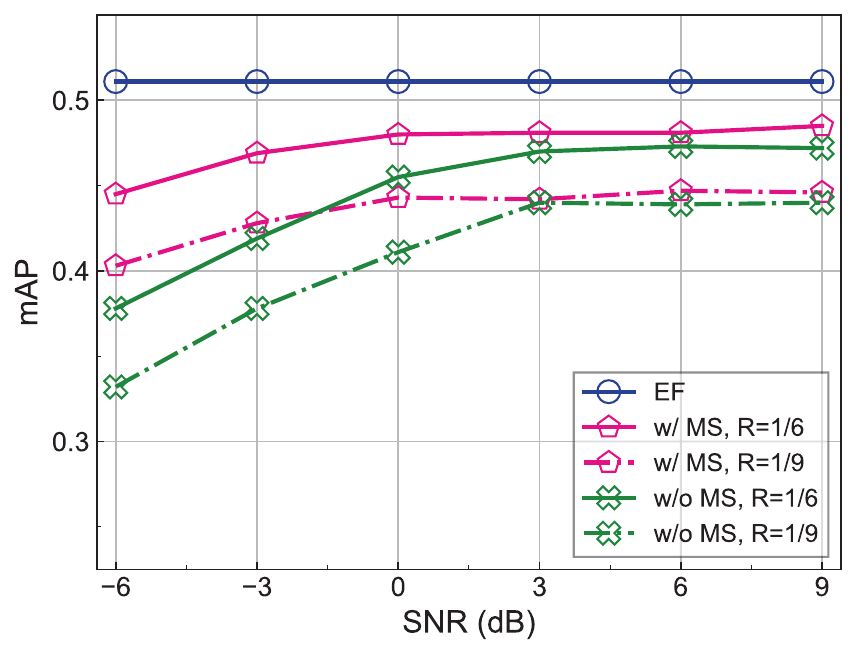}
      \label{cr6_awgn_ms_ablation}}
    \hspace{0.25in}
    \subfloat[Rayleigh channel.]{
      \includegraphics[width=3.3 in]{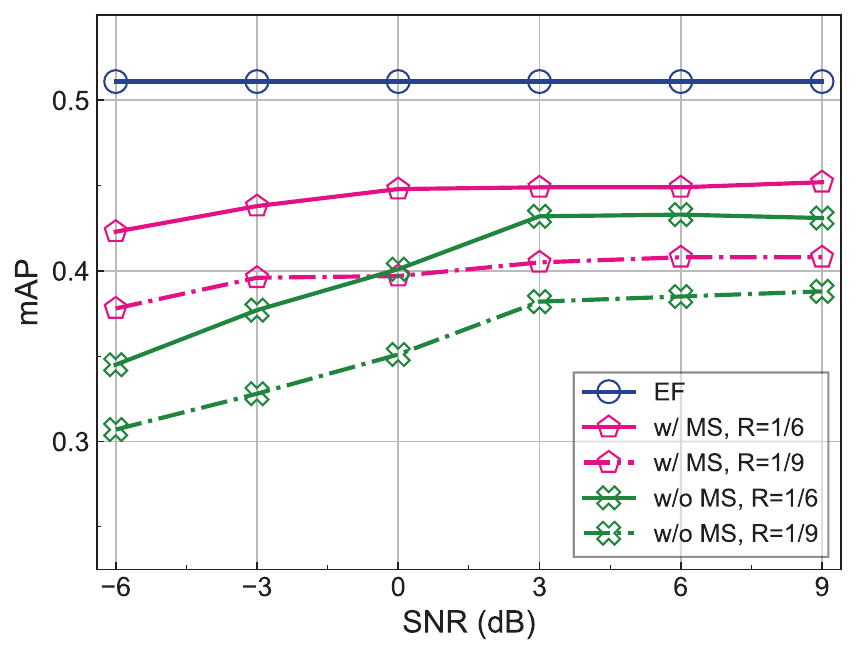}
      \label{cr6_rayleigh_ms_ablation}}
    \caption{mAP versus SNR for proposed system with and without the multi-scle codec (MS) under AWGN and Rayleigh channels.}
    \label{cr6_ms_ablation}
  \end{figure*}

  \begin{figure*}[h]
    \centering
    \subfloat[Confusion matrix of the proposed system without KG.]{
      \includegraphics[width=3.5 in]{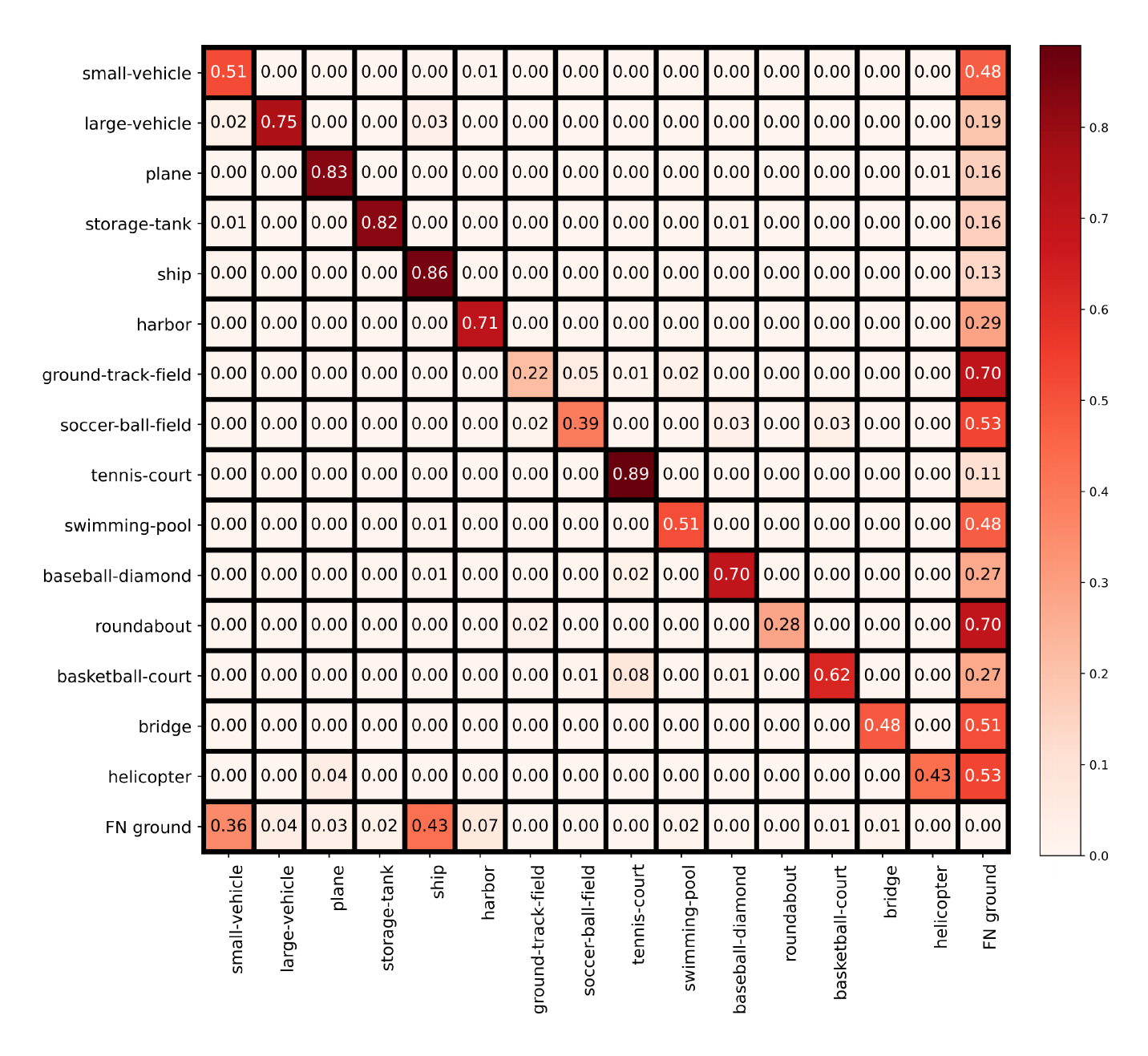}
      \label{confusion_mat_without_KG}}
    \subfloat[Confusion matrix of the proposed system with KG.]{
      \includegraphics[width=3.5 in]{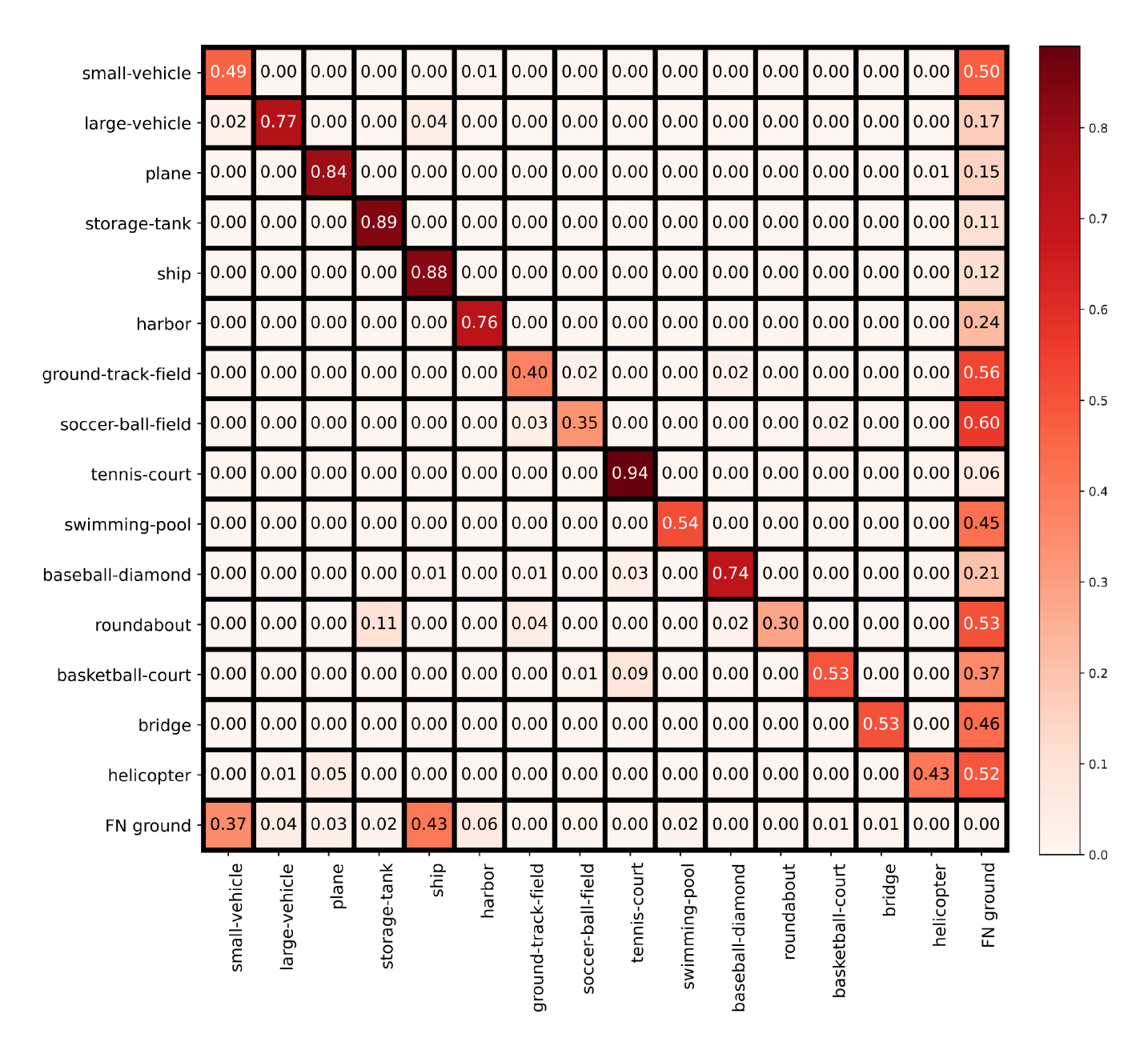}
      \label{confusion_mat_with_KG}}
    \caption{Confusion matrices of the proposed system across all fifteen categories. The label of the horizontal axis and vertical axis represents the ground truth and prediction, respectively. The confusion matrices are calculated under AWGN channel. (R $=$ 1/6, SNR $=$ 0dB).}
    \label{confusion_mat}
  \end{figure*}

  \begin{figure}[h]
    \centering
    \includegraphics[width=3.5 in]{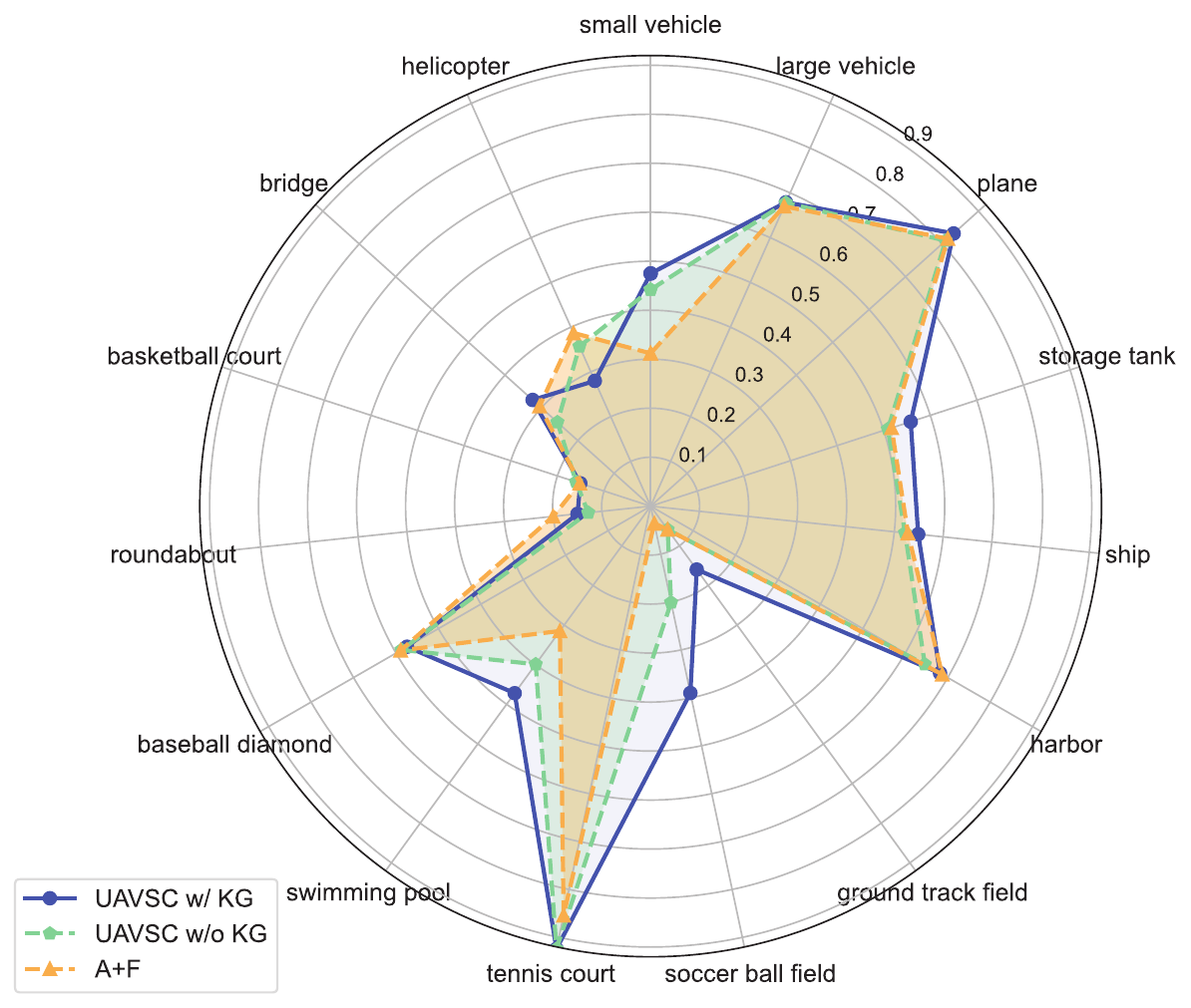}
    \caption{AP across all fifteen categories under AWGN channel. (R $=$ 1/6, SNR $=$ 0dB).}
    \label{cr6_awgn_snr0_radar}
  \end{figure}

  Table \ref{apcomprision} illustrates the outcomes of detailed evaluations of the systems under specific conditions. The performance enhancement or degradation of the proposed system across each metric is benchmarked against the A+F under identical conditions. Generally, the proposed system significantly elevates object detection precision, with improvements becoming more significant as the channel conditions worsen. Most improvements are astonishingly above 50\% in the Rayleigh channel with SNR $=$ -6dB. Considering mAP at different IoU thresholds, the proposed system consistently exhibits a substantial improvement in AP$_{50}$ compared to AP$_{50:95}$ and AP$_{75}$. The reason is that the compression-recovery scheme can still maintain the approximate contour of the object during image recovery, resulting in accurate localization under precise IoU thresholds. Therefore, the extent to which the proposed method can be improved is limited. Based on the AP of different scales, it can be revealed that the proposed system improves significantly on small-scale objects in most cases. This improvement can be attributed to the utilization of FPN and the multi-scale codec. The multi-scale semantic features encapsulate rich semantic features from high-level features, along with high-resolution and location information from the underlying semantic features.

  Fig. \ref{cr6_awgn_SAmodule_ablation} and Fig. \ref{cr6_rayleigh_SAmodule_ablation} present the mAP of the proposed system with and without the SA module under AWGN and Rayleigh channels, respectively. To underscore the effectiveness of the SA module, SNRs of -6dB, 0dB, and 9dB are selected to train the systems without the SA module, which are subsequently tested across all SNRs. Typically, the system without the SA module trained at a specific SNR experiences precision degradation at other SNRs. Specifically, the system without the SA module trained at SNRs of -6dB and 9dB encounters challenges in performing the detection task at SNRs of 9dB and -6dB. The system without the SA module trained at SNR $=$ 0dB is smoother in the face of both low and high SNRs, but the precision is still degraded. The observation from Fig. \ref{cr6_awgn_SAmodule_ablation} is noteworthy, where the system without the SA module trained at SNR $=$ 0dB performs better at SNR $=$ 9dB than at SNR $=$ 0dB under the AWGN channel. However, it exhibits suboptimal performance at SNRs of 3dB and 6dB. This anomaly indicates uncertainty in the performance of the system trained at a fixed SNR. In comparison, the system with the SA module eliminates the requirement for training at a specific SNR, presenting an all-in-one advantage and demonstrating a smoother trend overall SNRs. Furthermore, the system with the SA module outperforms all systems without that. Therefore, the incorporation of the SA module not only resolves the challenge of frequent model switching in a variable communication environment but also marginally enhances precision through the attention mechanism of feature maps.

  Fig. \ref{cr6_kg_ablation} demonstrates the ablation study performance of the proposed system with and without the knowledge graph. Under the AWGN channel, integrating the knowledge graph achieves an average boost of 2.94\% and 1.18\% at two bandwidth compression ratios, respectively. In terms of the Rayleigh channel, the improvements are 2.97\% and 0.9\%. The mAP curves for both channels exhibit similar trends. Concretely, the improvement from integrating the knowledge graph is more pronounced at R $=$ 1/6, while it is less noticeable at R $=$ 1/9. Moreover, the improvement is higher under low SNRs than high SNRs while R $=$ 1/9. The conclusion can be drawn that the benefits from integrating the knowledge graph become more significant as noise intensifies, and the benefits diminish as the bandwidth compression ratio increases. The primary reason for this is that the impact of compression distortion on semantic features is more significant than the interference caused by noise.

  Fig. \ref{cr6_ms_ablation} demonstrates the results of an ablation study assessing the impact of a multi-scale codec on mAP under varying SNR in both AWGN and Rayleigh channel conditions. The experiment compares systems with and without the multi-scale codec, with the latter utilizing spliced semantic features across scales for compression and encoding. In the AWGN channel, at bandwidth compression ratios of 6 and 9 with SNR $\leq$ 0, the average improvements are 4.73\% and 5.1\%, respectively. In the Rayleigh channel, the average improvements are 6.2\% and 6.16\%. The results reveal that the multi-scale codec significantly enhances mAP, particularly at low SNRs, indicating improved robustness in communication systems. The system without the codec shows stability at high SNRs. However, the performance gap persists, highlighting the substantial benefits of integrating multi-scale technology.

  \begin{figure*}[t]
    \centering
    \includegraphics[width=6 in]{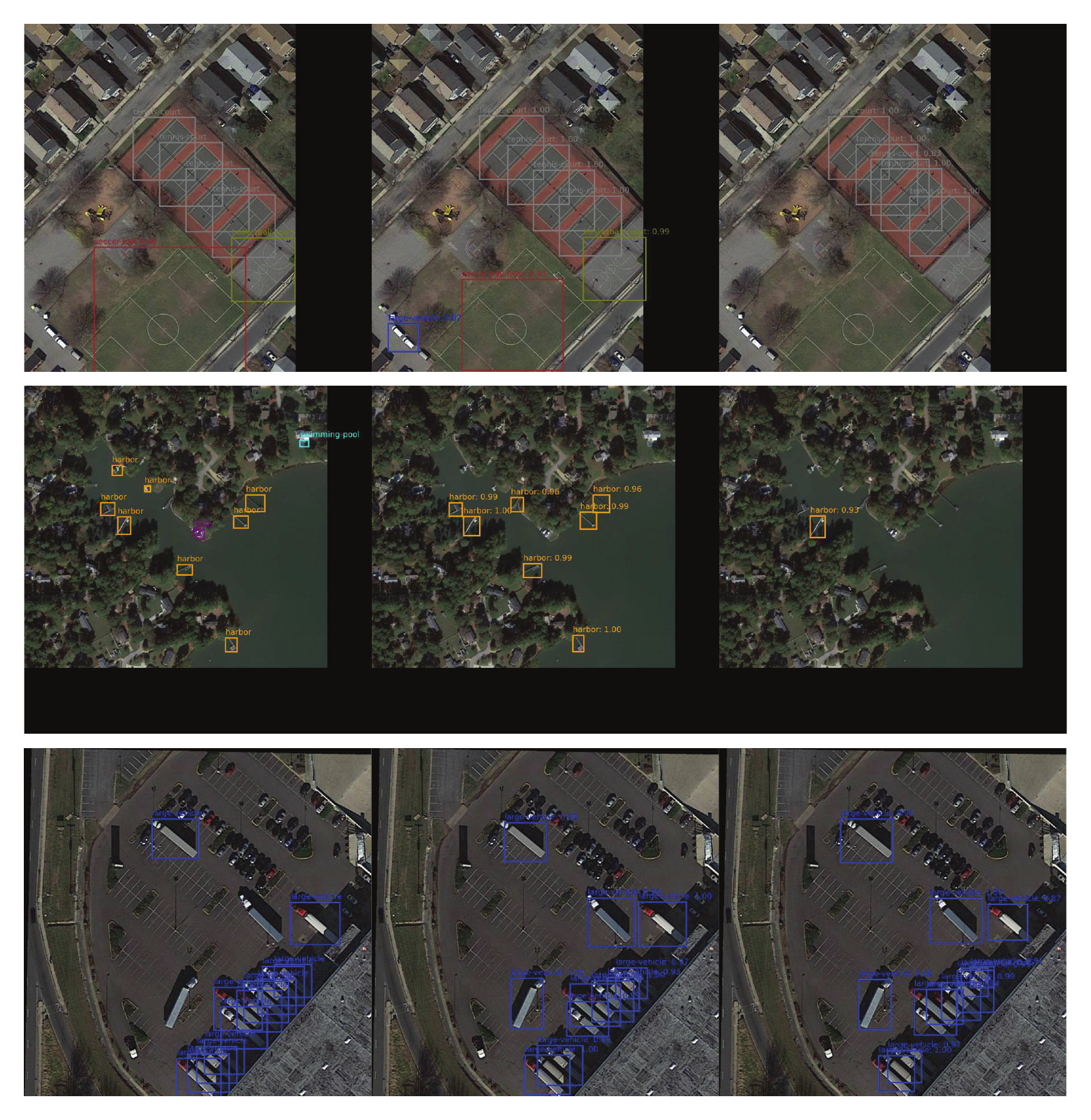}
    \caption{Visual comparison between the proposed system and A+F (AWGN channel, R = 1/6, SNR = -3dB). The first column presents the ground truth, the second column displays the results from the proposed system, and the third column shows the results obtained using the A+F system.}
    \label{vision}
  \end{figure*}

  In order to further analyze the details of how the knowledge graph can improve performance, we conduct a statistical analysis of the reasons for each category error and present it in the form of two confusion matrices. As depicted in Fig. \ref{confusion_mat}, the two confusion matrices represent the statistics of the proposed system with and without KG, respectively. Specifically, the horizontal axis label represents the ground truth, the vertical axis label represents the predicted value and the colour from light to dark red represents the degree of impact on the corresponding category hierarchically. Taking the last item in the first row as an example, for the \emph{small vehicle} category, 0.40 means that 25\% of the predictions are correct, 0.01 means that 1\% of the predictions are misclassified as \emph{harbor} category, and 0.5 means that 50\% of the prediction errors come from not being located by the detector. From Fig. \ref{confusion_mat}, we can see that almost all categories of prediction errors are caused by inaccurate localization. The problem is improved after integrating the knowledge graph. The prediction accuracy of most categories is also promoted. However, the misclassification rate in some cases increases because the detector integrated with the knowledge graph is excessively sensitive. For instance, the proportion of the \emph{roundabout} category being misclassified as \emph{storage tank} and \emph{ground track field} increased by 11\% and 2\%. This is a limitation of our work and need to be resolved and analyzed in future work.

  In Fig. \ref{cr6_awgn_snr0_radar}, the APs for all categories are presented under the AWGN channel while R $=$ 1/6 and SNR $=$ 0dB. It can be observed that there is a significant difference in the APs among different categories, which is partly due to the imbalance of instance samples. Large objects such as \emph{large vehicle} and \emph{plane} are more likely to be detected accurately. However, detecting small objects is extremely challenging. The challenge in detecting small vehicles lies in their small size, making them more difficult to detect after compression and noise interference. Detecting sports fields is also challenging due to the similarity of their texture features to the surrounding irrelevant background. According to Fig. \ref{cr6_awgn_snr0_radar}, the detection precision is promoted for almost all categories with the proposed system compared to the A+F. Considering the knowledge graph, it can be seen that integrating it has enhanced the precision for some small-scale objects such as \emph{small vehicle}, \emph{storage tank}, and some sport field objects, like \emph{ground track field}.


  \begin{table}[b]
    \renewcommand{\arraystretch}{1.5}
    \centering
    \caption{The statistics on efficiency factors of the UAV.}
    \begin{tabular}{cccc}
    \hline
    \textbf{Method} & \textbf{Parameters} & \textbf{GFLOPs} & \textbf{FPS} \\
    \hline
    UAVSC & $13.79M$ & $40.5$ & $239$ \\
    A+F & $17.09M$ & $131.7$ & $225$ \\
    \hline
    \end{tabular}
    \label{parameters}
  \end{table}

  We present a visual comparison between the proposed system and A+F for the sample image of the dataset in Fig. \ref{vision}. Note that the first column presents the ground truth, the second column displays the results from the proposed system, and the third column shows the results obtained using the A+F system. The results demonstrate a significant reduction in missed detections with our approach, highlighting its effectiveness in identifying objects that the A+F failed to detect. Furthermore, the proposed system exhibits a distinct advantage in detecting small objects. The results also indicate that the proposed system operates with higher confidence levels, further underscoring its efficacy in accurately identifying objects across varying conditions.

  The statistics on some efficiency factors on the UAV are shown in Table \ref{parameters}. The proposed system exhibits apparent advantages. It can be seen that the number of parameters in the proposed system is smaller, with 19.3\% less compared to A+F. This compactness contributes to reduced storage and transmission overheads. The computation requirement of the proposed system is also much lower than A+F, which is as high as an astonishing 131.7 GFLOPs. It implies that the proposed system can operate more efficiently within limited computation resources, making it well-suited for UAV communication scenarios. The proposed system achieves a higher FPS of 239, surpassing the 225 FPS of A+F. It suggests that the proposed system can process image information more rapidly in real-time scenarios, allowing UAVs to make timely decisions and respond to environmental changes. Therefore, considering the advantages of the proposed system, it emerges as a superior computational model in UAV communication scenarios.

\section{Conclusion}
  A UAV cognitive semantic communication system was proposed for object detection by exploiting knowledge graph. Moreover, a multi-scale codec was developed for semantic compression to reduce data transmission while ensuring detection accuracy. Considering the complexity and dynamicity of UAV communication scenarios, an SNR adaptive module with robust channel adaptation capability was introduced. Furthermore, a detection scheme was proposed by exploiting auxiliary knowledge of the knowledge graph to overcome channel noise interference and compression distortion. Simulation results demonstrated that the proposed system has superior detection accuracy, communication robustness and computation efficacy under low bandwidth compression ratios and low SNR regimes.

\bibliographystyle{IEEEtran}
\bibliography{reference}

\vspace{-12mm}

\begin{IEEEbiography}[{\includegraphics[width=1.0in,height=1.15in,clip,keepaspectratio]{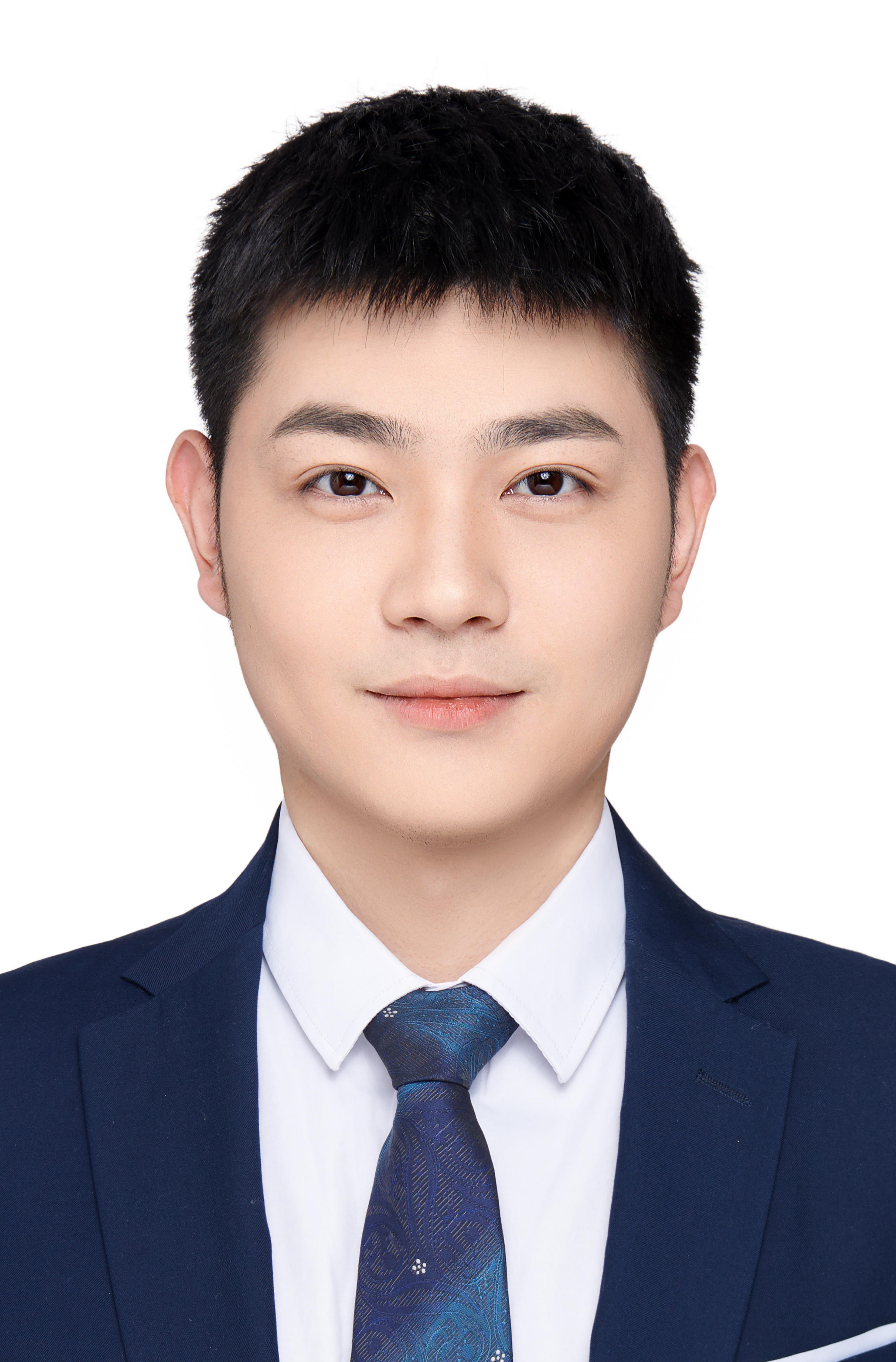}}]
  {Xi Song} received the B.Eng. degree in software engineering from Shandong Jianzhu University, Jinan, China. He is currently working toward the M.Eng. degree in information and communication engineering with Nanjing University of Aeronautics and Astronautics, Nanjing, China. His research interests include semantic communication, knowledge graph, and deep learning.
\end{IEEEbiography}

\vspace{-5mm}
\begin{IEEEbiography}[{\includegraphics[width=1.0in,height=1.15in,clip,keepaspectratio]{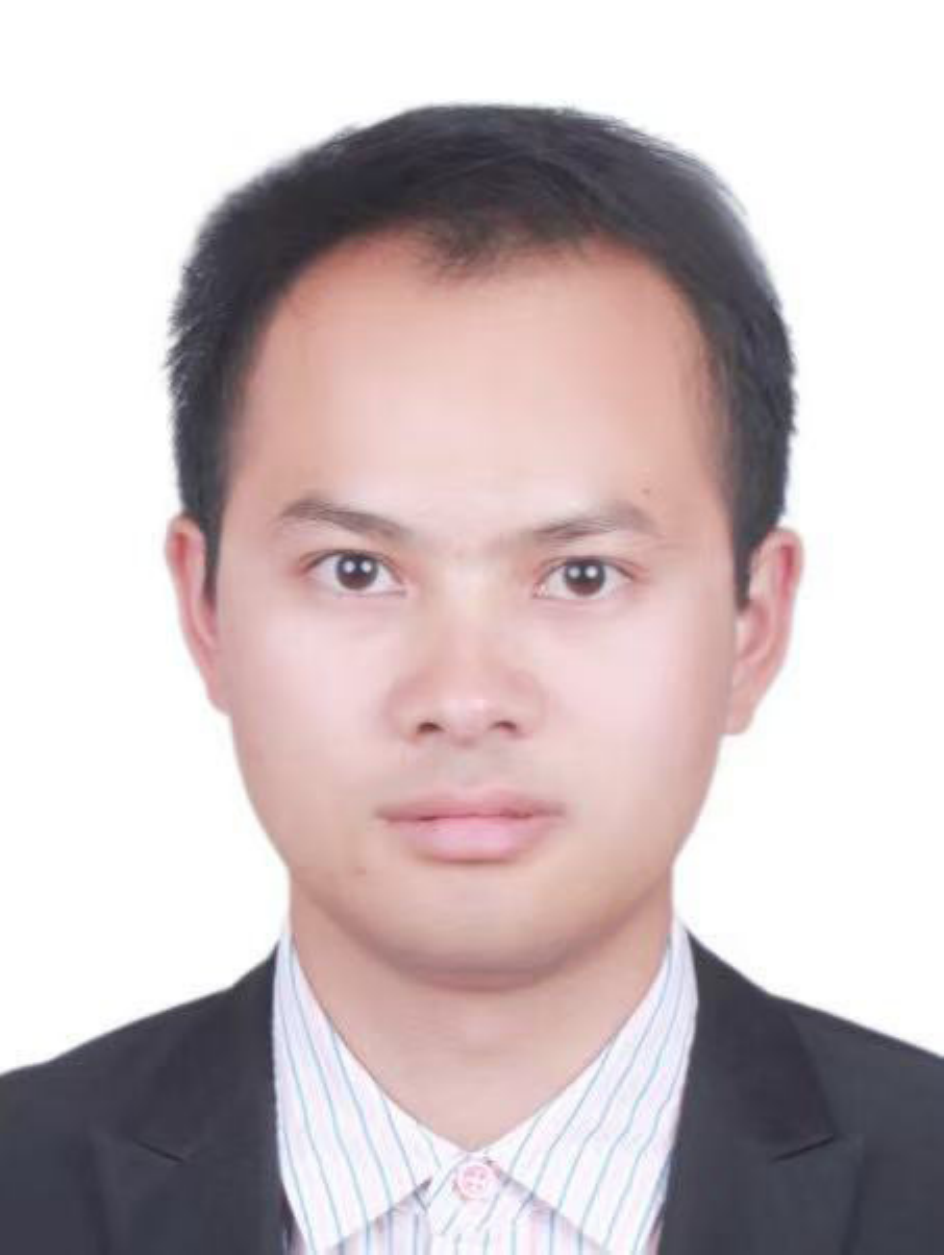}}]
  {Fuhui Zhou} (Senior Member, IEEE) is currently a Full Professor with the College of Artificial Intelligence, Nanjing University of Aeronautics and Astronautics, Nanjing, China, where he is also with the Key Laboratory of Dynamic Cognitive System of Electromagnetic Spectrum Space. His research interests include cognitive radio, cognitive intelligence, knowledge graph, edge computing, and resource allocation. 

  Prof. Zhou has published over 200 papers in internationally renowned journals and conferences in the field of communications. He has been selected for 1 ESI hot paper and 13 ESI highly cited papers. He has received 4 Best Paper Awards at international conferences such as IEEE Globecom and IEEE ICC. He was awarded as 2021 Most Cited Chinese Researchers by Elsevier, Stanford World’s Top 2\% Scientists, IEEE ComSoc Asia-Pacific Outstanding Young Researcher and Young Elite Scientist Award of China and URSI GASS Young Scientist. He serves as an Editor of IEEE Transactions on communication, IEEE Systems Journal, IEEE Wireless Communication Letters, IEEE Access and Physical Communications.
\end{IEEEbiography}

\vspace{-5mm}
\begin{IEEEbiography}[{\includegraphics[width=1in,height=1.15in,clip,keepaspectratio]{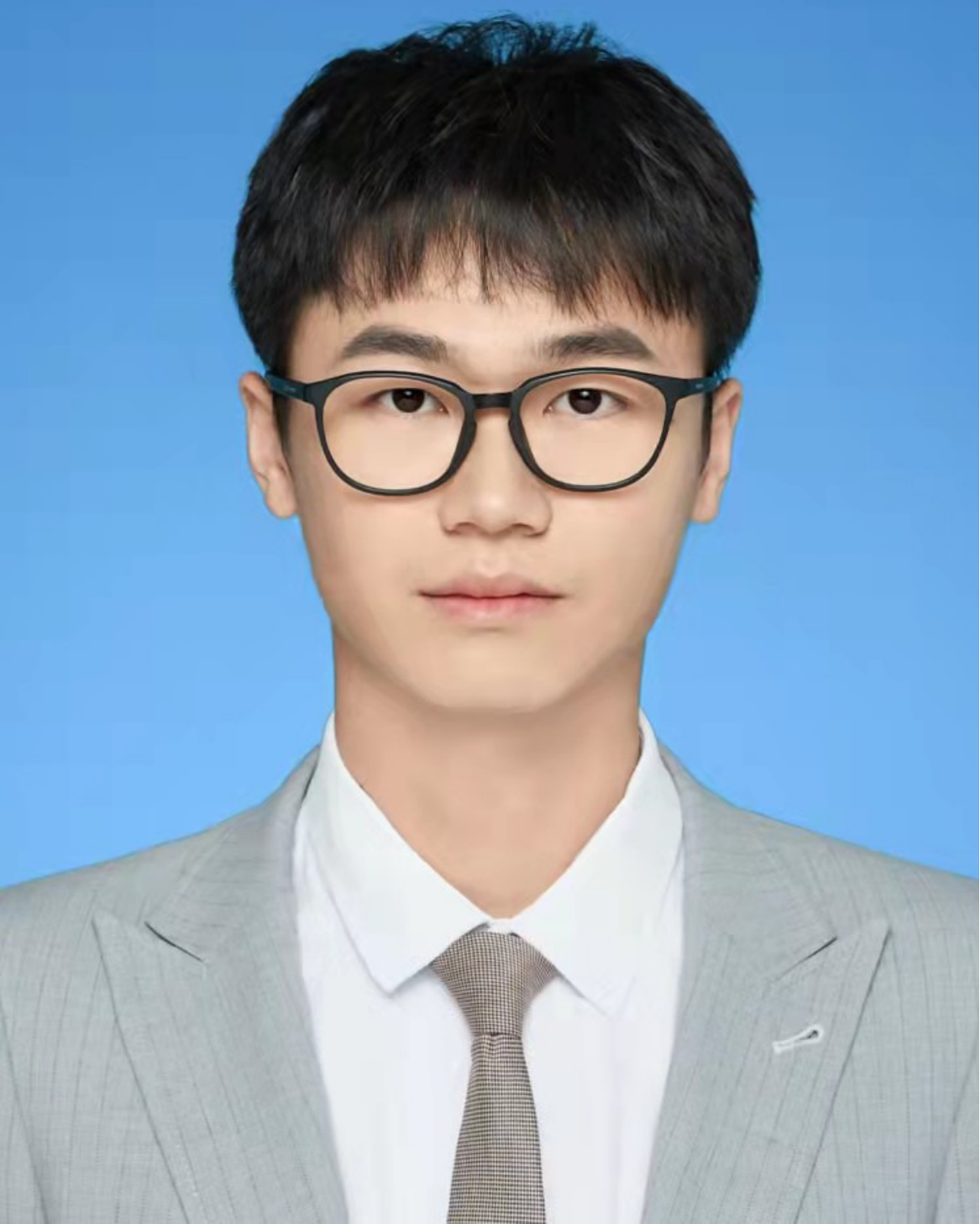}}]
  {Rui Ding} is currently working toward the Ph.D. degree with the School of Electronic and Information Engineering, Nanjing University of Aeronautics and Astronautics, Nanjing, China. His research interests include deep reinforcement learning, deep learning, and deep active inference with applications in resource allocation, spectrum management for wireless communications networks, including data-and-knowledge driven cognitive and decision-making, intelligent spectrum sharing, and UAV communications.
\end{IEEEbiography}

\vspace{-5mm}
\begin{IEEEbiography}[{\includegraphics[width=1in,height=1.15in,clip,keepaspectratio]{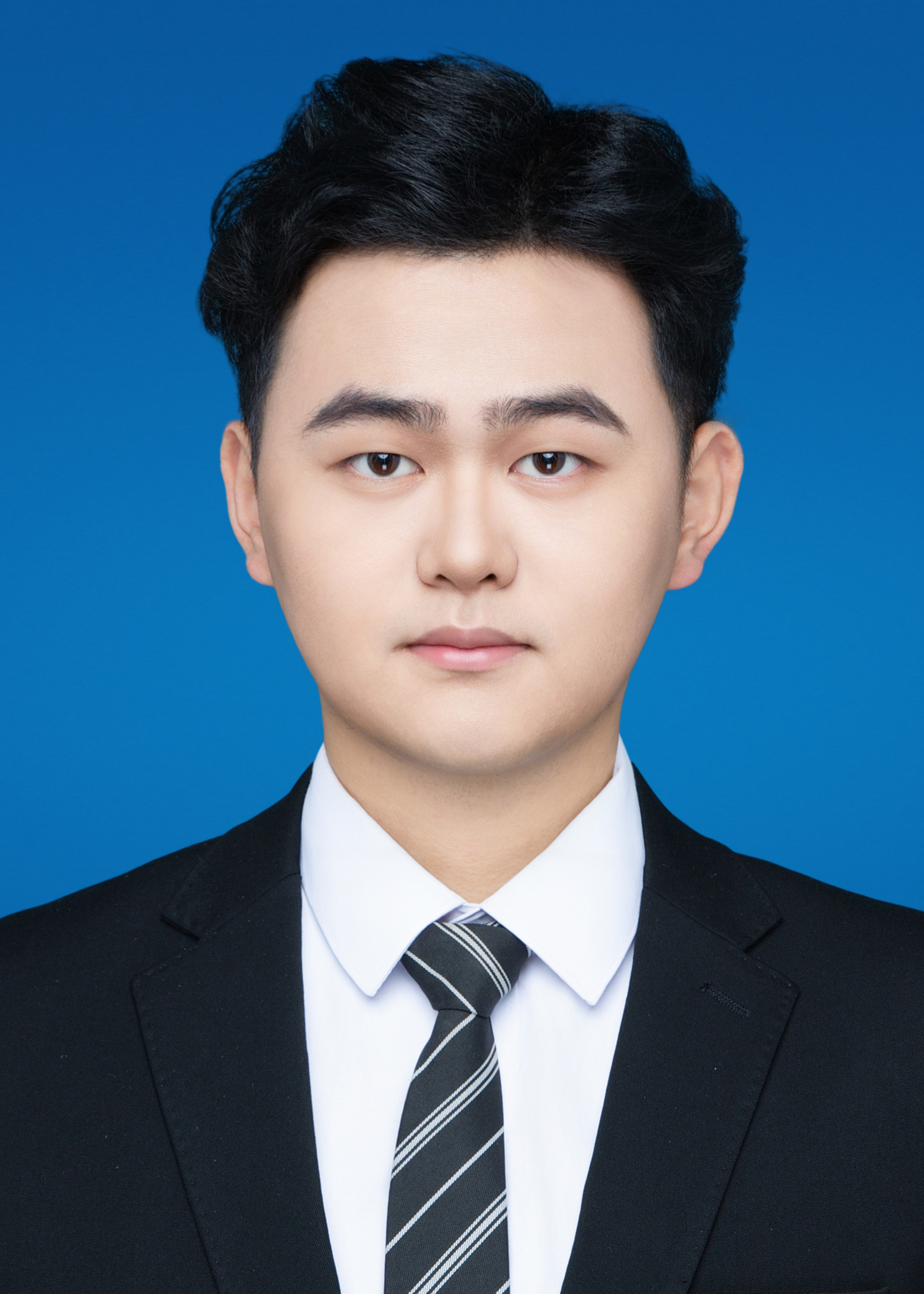}}]
  {Zhibo Qu} received the B.Eng. degree in communication engineering from Shandong University, Qingdao, China. He is currently working toward the M.Eng. degree in electronic and information engineering from Nanjing University of Aeronautics and Astronautics, Nanjing, China. His research interests include knowledge graph, natural language processing, and deep learning.
\end{IEEEbiography}

\begin{IEEEbiography}[{\includegraphics[width=1in,height=1.15in,clip,keepaspectratio]{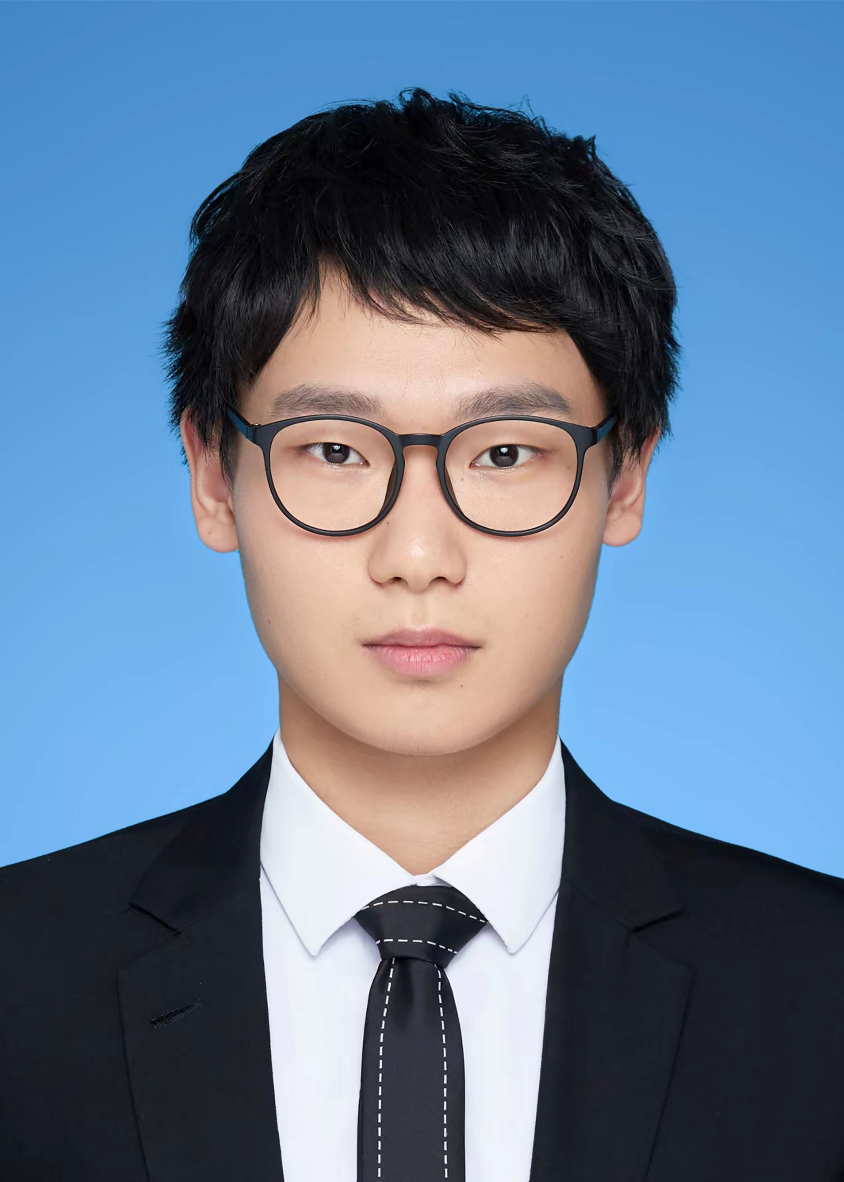}}]
  {Yihao Li} received the M. Eng. degree in Information and Communication Engineering from the Nanjing University of Aeronautics and Astronautics in 2024 and is currently pursuing the PhD degree in School of Information Science and Engineering from the Southeast University. His research interests include semantic communication, knowledge graph, natural language processing, and deep learning.
\end{IEEEbiography}

\vspace{-5mm}
\begin{IEEEbiography}[{\includegraphics[width=1in,height=1.15in,clip,keepaspectratio]{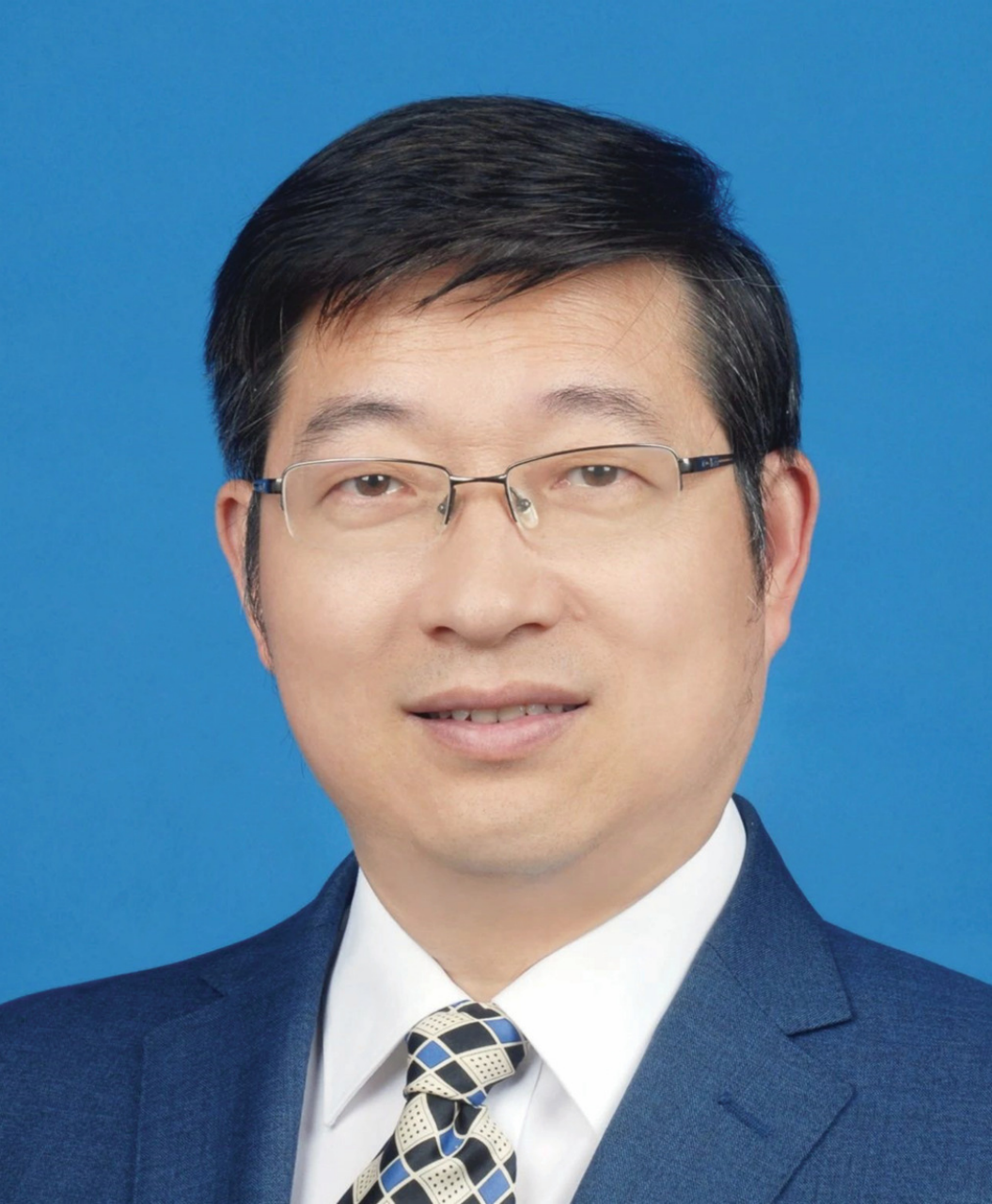}}]
  {Qihui Wu} (Fellow, IEEE) received the B.S. degree in communications engineering, the M.S. and Ph.D. degrees in communications and information systems from the Institute of Communications Engineering, Nanjing, China, in 1994, 1997, and 2000, respectively. From 2003 to 2005, he was a Postdoctoral Research Associate with Southeast University, Nanjing, China. From 2005 to 2007, he was an Associate Professor with the College of Communications Engineering, PLA University of Science and Technology, Nanjing, China, where he was a Full Professor from 2008 to 2016. SinceMay 2016, he has been a Full Professor with the College of Electronic and Information Engineering, Nanjing University of Aeronautics and Astronautics, Nanjing, China. From March 2011 to September 2011, he was an Advanced Visiting Scholar with the Stevens Institute of Technology, Hoboken, USA. His current research interests span the areas of wireless communications and statistical signal processing, with emphasis on system design of software defined radio, cognitive radio, and smart radio.
\end{IEEEbiography}

\vspace{-5mm}
\begin{IEEEbiography}[{\includegraphics[width=1in,height=1.15in,clip,keepaspectratio]{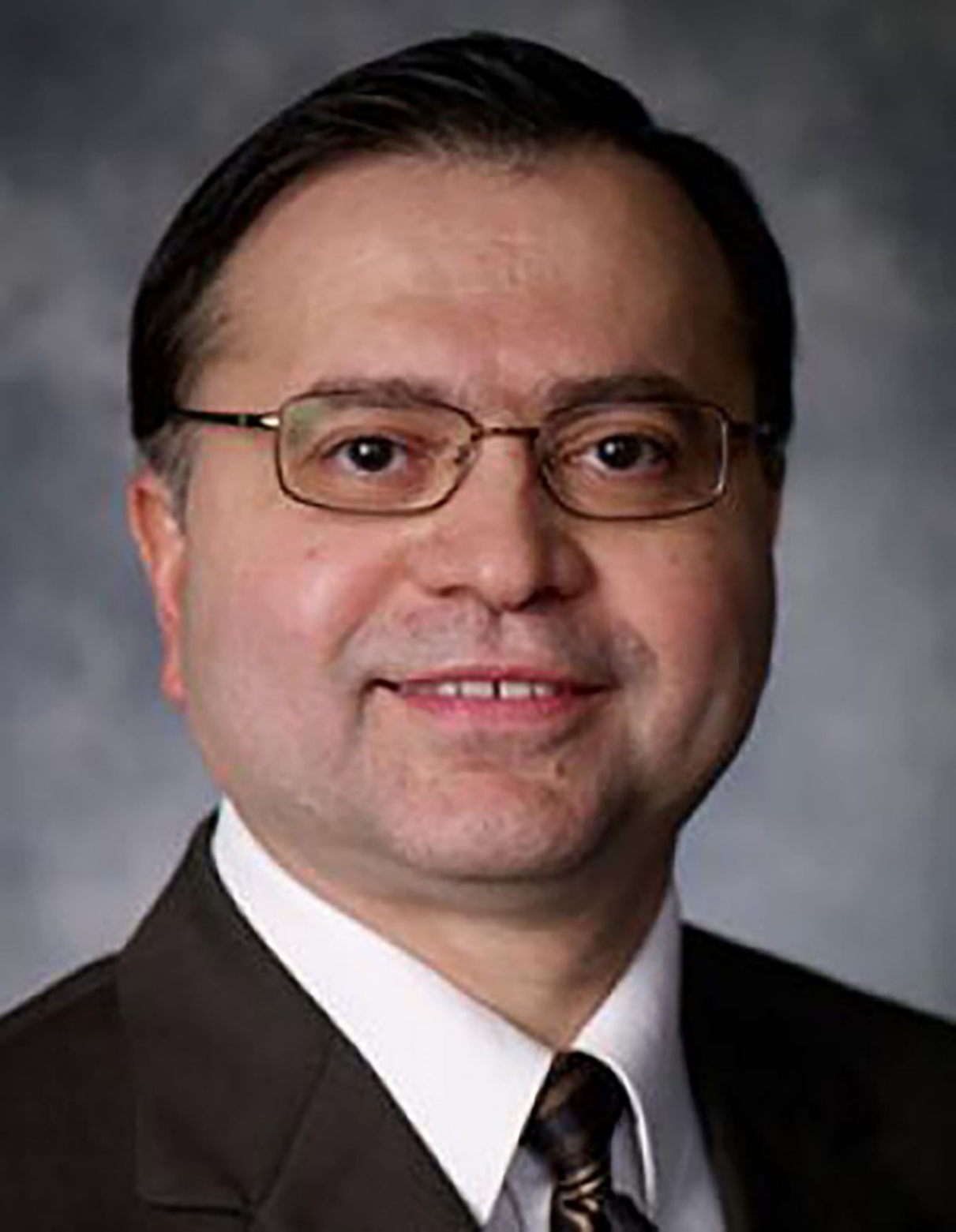}}]
  {Naofal Al-Dhahir} (Fellow, IEEE) received the Ph.D. degree from Stanford University. He is currently an Erik Jonsson Distinguished Professor and the ECE Associate Head of UT-Dallas. He was a Principal Member of Technical Staff with the GE Research Center and AT\&T Shannon Laboratory from 1994 to 2003. He is a co-inventor of 43 issued patents, the co-author of over 600 articles, and corecipient of eight IEEE best paper awards. He is a fellow of AAIA and U.S. National Academy of Inventors and a member of European Academy of Sciences and Arts. He received 2019 IEEE COMSOC SPCC Technical Recognition Award, 2021 Qualcomm Faculty Award, and 2022 IEEE COMSOC RCC Technical Recognition Award. He served as the Editor-in-Chief for IEEE Transactions on Communications from January 2016 to December 2019.
\end{IEEEbiography}

\end{document}